\documentclass[nofootinbib]{revtex4}
\usepackage{graphicx}
\usepackage{dcolumn}
\usepackage{amsmath,amssymb,epsfig}
\usepackage{paralist}
\usepackage{comment}
\usepackage{graphicx}
\usepackage{multirow}
\usepackage{color,soul}
\usepackage{suffix}
\usepackage{mathtools}

\allowdisplaybreaks

\renewcommand{\vec}[1]{\boldsymbol{\mathrm{#1}}}

\begin{document}

\title{Wave-optical study of the Einstein cross formed by a quadrupole gravitational lens}

\author{Slava G. Turyshev$^{1}$, Viktor T. Toth$^2$}

\affiliation{\vskip 3pt
$^1$Jet Propulsion Laboratory, California Institute of Technology,\\
4800 Oak Grove Drive, Pasadena, CA 91109-0899, USA}

\affiliation{\vskip 3pt
$^2$Ottawa, Ontario K1N 9H5, Canada}

\date{\today}

\begin{abstract}

We study imaging of point sources with a quadrupole gravitational lens while focusing on the formation and evolution of the  Einstein cross formed on the image sensor of an imaging telescope. For this, we consider the optical properties of an oblate gravitational lens that is characterized, in addition to a monopole potential, by the presence of a quadrupole zonal harmonic. We use a new type of a diffraction integral that we developed to study generic, opaque, weakly aspherical gravitational lenses. To  evaluate this integral, we use the method of stationary phase that yields a quartic equation with respect to a Cartesian projection of the observer's position vector with respect to the vector of the impact parameter. The resulting quartic equation can be solved analytically using the method first published by Cardano in 1545. We find that the resulting solution provides a good approximation of the electromagnetic (EM)  field almost everywhere in the image plane, yielding the well-known astroid caustic of the quadrupole lens. The sole exception is the immediate vicinity of the caustic boundary, where a numerical treatment of the diffraction integral yields better results. We also convolve the quartic solution for the EM field on the image plane with the point-spread function of a thin lens imaging telescope. By doing so, we are able to explore the direct relationship between the algebraic properties of the quartic solution for the EM field, the geometry of the astroid caustic, and the geometry and shape of the resulting Einstein-cross that appear on the image plane of the thin lens telescope. The new quartic solution leads to significant improvements in numerical modeling as evaluation of this solution is computationally far less expensive than a direct numerical treatment of the new diffraction integral. In the case of the solar gravitational lens (SGL), the new results drastically improve the speed of numerical simulations related to sensitivity analysis performed in the context of high-resolution imaging of exoplanets.

\end{abstract}


\maketitle

\section{Introduction}

Recently, while studying the solar gravitational lens (SGL), we developed an approach that may be used to treat gravitational lensing by generic extended gravitating bodies, see details in \cite{Turyshev-Toth:2021-multipoles}.  For that, we considered  light propagation within the wave-optical treatment and solved the Mie problem while working within the first post-Newtonian approximation of the general theory of relativity. To characterize physical properties of a large, rotating, gravitating object,  we treated the axisymmetric gravitational potential of the body by modeling it in the form of in infinite sum of zonal harmonic terms. This work provided us with a wave-theoretical description of the optical properties of extended axisymmetric gravitational lenses that can be applied to treat many realistic astrophysical lenses.

We used the new wave-optical treatment to study gravitational lenses with increasing complexity.  Specifically, in \cite{Turyshev-Toth:2021-multipoles} we have developed a wave-optical treatment of gravitational lensing on compact lenses possessing an arbitrary number of gravitational multipoles. In \cite{Turyshev-Toth:2021-caustics} we studied the propagation of light in the extended solar gravitational field and examined the caustics formed in an image plane in the strong interference region. We investigate the EM field in several important regions, namely i) the area in the inner part of the caustic and close to the optical axis, ii) the region outside the caustic, and iii) the region in the immediate vicinity of the caustic, especially around its cusps and folds.  We show that in the first two regions the physical behavior of the EM field may be understood using the method of stationary phase. The new method allows us to investigate the EM field in this important region, which is characterized by rapidly oscillating behavior. In \cite{Turyshev-Toth:2021-imaging}, we consider  gravitational field of a lens that is dominated by a quadrupole moment, which results in forming an astroid caustic on the image plane. We show that an imaging telescope placed inside the astroid caustic observes four bright spots, forming the well-known pattern of an Einstein cross.  If positioned outside the caustic, the telescope will see two bright spots that are well described by a monopole lens \cite{Turyshev-Toth:2017,Turyshev-Toth:2020-extend}. This transition is interesting motivating us to study the Einstein cross evolution in more details.

As often is the case with generic gravitation lensing, expressions characterizing the EM field in the image plane involve double integrals with rapidly oscillating integrands \cite{Schneider-Ehlers-Falco:1992}. In the case of weakly aspherical, optically opaque lenses such as our Sun, light from a distant source appears, to a very good approximation, in the form of an Einstein ring, as seen by an observer in the focal region of the lens. To treat these cases, we can introduce a cylindrical, lens-centric coordinate system and use the method of stationary phase in the radial direction. This amounts to finding the impact parameter that corresponds to the apparent radius of this Einstein ring.

In the angular direction, however, depending on the magnitude of the zonal harmonic contributions, light may appear in the form of extended arcs or even as a fully formed Einstein ring. Accordingly, the method of stationary phase in the angular direction may help us find the brightest spots of such arcs, but not their extended structure. Therefore, leaving this integral to be evaluated using numerical methods seems prudent. However, even in this case further analytical progress is possible, especially in the case when the quadrupole term that characterizes the gravitational potential dominates over the other harmonics.  In the case when only quadrupole distortion is present, an observer, depending on her position with respect to the astroid caustic formed by a quadrupole lens, would see the formation and evolution of the Einstein cross (see
\cite{Turyshev-Toth:2021-caustics} for discussion). This case is interesting and deserves study.

Our main objective here is to study impact of the astroid caustic formed in the point spread function (PSF) of a quadrupole gravitational lens. For that we aim to investigate the image formation process of such a lens that leads to forming an Einstein cross on the image plane. This paper is organized as follows:
In Section~\ref{sec:stat-ph}, we consider the case of a quadrupole gravitational lens. We use the method of stationary phase to find a solution to the integral in (\ref{eq:B2}), reducing the entire problem to solving a quartic equation.
In Section~\ref{sec:quart-inaging}, we consider imaging point sources with a quadrupole lens.
Finally, in Section ~\ref{sec:end}, we discuss results and outline prospective next steps.
In Appendix~\ref{sec:mono-quad-cases}, we work out the limits of the quartic solution far from the quadrupole caustic boundary.
Lastly, Appendix~\ref{sec:cos} presents an alternate form of the quartic equation, a result used in the main text to resolve trigonometric sign ambiguities.

\section{A quartic-based solution for the EM field}
\label{sec:stat-ph}

\subsection{Wave optical treatment of extended gravitatiomal lenses}

Our starting point is our previously developed solution that can be summarized as follows. We consider the propagation of a high-frequency EM wave (i.e., neglecting terms $\propto(kr)^{-1}$ and for $r\gg r_g$; here, $k$ is the wavenumber, $r_g$ is the Schwarzschild radius of a gravitational lens and $r$ is the distance from that lens) and derive the EM field in the strong interference region of the gravitational lens \cite{Turyshev-Toth:2021-multipoles}. We use a lens-centric coordinate system with its $z$-axis oriented along the wavevector $\vec{k}$ of the incident wave. We introduce the vector of the impact parameter, $\vec b$, and define the image plane, located at distance $z$ from the lens, where it projects light from the source to form an image. Locations in this image plane are marked using $\vec x$. Also, we introduce the unit vector in the direction of the rotation axis of the lens as $\vec s$:
\begin{eqnarray}
{\vec b}&=&b(\cos\phi_\xi,\sin \phi_\xi,0)=b \vec n_\xi, \\
{\vec x}&=&\rho(\cos\phi,\sin \phi,0)=\rho \vec n,\\
{\vec s}&=&(\sin\beta_s\cos\phi_s,\sin\beta_s\sin\phi_s,\cos\beta_s).
\label{eq:note}
\end{eqnarray}

In this geometry, up to terms of ${\cal O}(r_g^2, \rho^2/z^2)$, the EM field in the image plane takes the following form (for derivation and details, see \cite{Turyshev-Toth:2021-multipoles} and references therein):
{}
\begin{eqnarray}
    \left( \begin{aligned}
{E}_\rho& \\
{H}_\rho& \\
  \end{aligned} \right) =\left( \begin{aligned}
{H}_\phi& \\
-{E}_\phi& \\
  \end{aligned} \right) &=&
E_0 \sqrt{2\pi kr_g}e^{i\sigma_0}B(\vec x)
  e^{i(kz -\omega t)}
 \left( \begin{aligned}
 \cos\phi& \\
 \sin\phi& \\
  \end{aligned} \right),
  \label{eq:DB-sol-rho}
\end{eqnarray}
with the remaining EM field components being negligibly small of the order of  $({E}_z, {H}_z)= {\cal O}({\rho}/{z})$ and where $\sigma_0$ is the constant $\sigma_0=-kr_g\ln kr_g/e-{\textstyle\frac{\pi}{4}}$ (see  \cite{Turyshev-Toth:2017,Turyshev-Toth:2019,Turyshev-Toth:2021-multipoles,Turyshev-Toth:2021-caustics} for details). The quantity $B(\vec x)$  is the complex amplitude of the EM field, given as
{}
\begin{eqnarray}
B(\vec x) &=&
\frac{1}{2\pi}\int_0^{2\pi} d\phi_\xi \exp\Big[-ik\Big(\sqrt{\frac{2r_g}{r}}\rho\cos(\phi_\xi-\phi)+
2r_g\sum_{n=2}^\infty \frac{J_n}{n} \Big(\frac{R_\odot }{\sqrt{2r_gr}}\Big)^n\sin^n\beta_s\cos[n(\phi_\xi-\phi_s)]\Big)\Big],
  \label{eq:B2}
\end{eqnarray}
where $J_n$ are the zonal harmonics characterizing the external gravitational potential of an axisymmetric lens.

The complex amplitude (\ref{eq:B2}) determines the properties of the EM field in the image plane positioned in the strong interference region of the lens. The quantity ${\rm PSF}=B(\vec x) B^*(\vec x) $, where the asterisk marks the complex conjugate, determines the energy density of the EM wave in the image plane produced, i.e.,  the point spread function (PSF) of the lens \cite{Turyshev-Toth:2021-multipoles,Turyshev-Toth:2021-caustics}. If the contribution of multipole moments (i.e., if $J_n=0, \forall n\geq 2)$ may be neglected, the integral (\ref{eq:B2}) reduces to $J_0(k\sqrt{2r_g/r}\rho)$, where $J_0$ is the Bessel-function of the first kind (see \cite{Turyshev-Toth:2017} and Appendix~\ref{sec:mono-case}). In a more general case, characterizing an axisymmetric lens using zonal harmonics, the diffraction integral (\ref{eq:B2})  is new. It cannot be reduced to any of the known mathematical functions or any of the known canonical integrals describing diffraction catastrophes \cite{Abramovitz-Stegun:1965,Berry-Howles:2010}.

This result, given by (\ref{eq:DB-sol-rho})--(\ref{eq:B2}), describes the EM field deposited in the image plane in the strong interference region of the lens (in the immediate vicinity of the optical axis, i.e., $\rho\sim r_g$) after a plane EM wave traveled through the vicinity of an extended, axisymmetric, and weakly aspherical gravitational lens of an optically opaque body. This solution was explored in \cite{Turyshev-Toth:2021-caustics,Turyshev-Toth:2021-imaging} where it was shown to produce caustics of different orders. In fact, each of the zonal harmonics, $J_n$, is responsible for producing a distinct set of caustics. In general, the structure and brightness of the resulting PSF is driven by all the harmonics $J_n$ present in (\ref{eq:B2}), each introducing its own spatial frequency (see \cite{Turyshev-Toth:2021-caustics} for discussion) that affects the variable rapid oscillations of the integrand, necessitating the use of numerical methods.

Our main current objective is the evaluation of the new diffraction integral  (\ref{eq:B2}) in the case of a quadrupole lens (i.e., when only $J_2$ is present)  and studying its  properties as far as imaging is concerned. Note that in the case of weakly aspherical extended gravitating bodies, such as the Sun, the gravitational phase shift in  the integrand of (\ref{eq:B2}) is typically dominated by the presence of two largest harmonics  - the monopole and quadrupole, $J_2$. Below, we will consider the case when $J_3$ and higher zonal harmonics can be safely ignored, i.e., the quadrupole lens.

\subsection{ Complex amplitude of the EM field for a quadrupole lens}

With only the $J_2$ zonal harmonic present, the complex amplitude of the EM field on the image plane, $B(\vec x)$,  from (\ref{eq:B2}), takes the form
{}
\begin{eqnarray}
B_2(\vec x)&=&\frac{1}{{2\pi}}\int_0^{2\pi} d\phi_\xi \exp\Big[-i\Big(\alpha\rho\cos(\phi_\xi-\phi)+\beta_2\cos[2(\phi_\xi-\phi_s)]\Big)\Big],
\label{eq:zer*1}
\end{eqnarray}
where we introduced the constants $\alpha$ and $\beta_2$ as
{}
\begin{eqnarray}
\alpha&=&k\sqrt\frac{2r_g}{r},~~~~\beta_2=kr_gJ_2 \Big(\frac{R_\odot }{\sqrt{2r_gr}}\Big)^2\sin^2\beta_s.
\label{eq:zerJ0}
\end{eqnarray}
Note that the integral (\ref{eq:zer*1})  (as well as (\ref{eq:B2})) is given in natural physical units, with parameters that have clear physical meaning. The parameter $\alpha$ represents the spatial frequency of the characteristic pattern that forms as a result of the quadrupole moment, while the parameter $\beta_2$ is the phase shift accumulated by the EM wave due to $J_2$ as it travels in the vicinity of the extended lens (see relevant discussion in \cite{Turyshev-Toth:2021-caustics}). As we discussed in \cite{Turyshev-Toth:2021-caustics},  the integral (\ref{eq:zer*1}) is responsible for forming an astroid caustic in the image plane, positioned in the strong interference region of the SGL.  This astroid caustic may be presented in a parametric form as
 {}
\begin{eqnarray}
x&=&\frac{4\beta_2}{\alpha}\cos^3\phi,
\label{eq:q_caust_j1}\qquad
y=\frac{4\beta_2}{\alpha}\sin^3\phi
\qquad \Rightarrow\qquad
\rho^2=\Big(\frac{4\beta_2}{\alpha}\Big)^2\Big(1-3\sin^2\phi\cos^2\phi\Big),
\label{eq:q_caust_j2}
\end{eqnarray}
with $\rho$ and $\phi$ being polar coordinates in the image plane.

To put our discussion in the context of the SGL, we consider a high-frequency EM wave (or, equivalently, a wave with short wavelength, $\lambda\sim 1\,\mu$m), and, using the Sun's value of $J_2=(2.25\pm0.09)\times 10^{-7}$  \cite{Park-etal:2017}, we estimate the values of the parameters $\alpha$ and $\beta_2$ defined by (\ref{eq:zerJ0}) as
{}
\begin{eqnarray}
\alpha&=&k\sqrt\frac{2r_g}{r}=48.97\,{\rm m}^{-1}\, \Big(\frac{1\,\mu{\rm m} }{\lambda}\Big)\Big(\frac{650{\rm AU} }{r}\Big)^\frac{1}{2},
\label{eq:zerJ}\\
\beta_2 &=&kr_gJ_2 \Big(\frac{R_\odot }{\sqrt{2r_gr}}\Big)^2\sin^2\beta_s=3518.34\, {\rm rad}\, \Big(\frac{J_2}{2.25\times 10^{-7}}\Big)\Big(\frac{1\,\mu{\rm m} }{\lambda}\Big)\Big(\frac{650{\rm AU} }{r}\Big)\sin^2\beta_s.
\label{eq:kbet-J2}
\end{eqnarray}
Using (\ref{eq:zerJ})--(\ref{eq:kbet-J2}), we estimate the scale of the astroid caustic given by (\ref{eq:q_caust_j2}) as ${4\beta_2}/{\alpha} =287.39\, {\rm m}\sqrt{{650\,{\rm AU} }/{r}}\sin^2\beta_s$. Therefore, for a given target, the size of the quadrupole caustic of the SGL is determined by the angle $\beta_s$ (determined by the sky position of the source), and the heliocentric distance to the image plane. The numerical values of the parameters $\alpha$ and $\beta_2$  (as well as their ratio ${4\beta_2}/{\alpha}$) characterize the challenge of evaluating integral (\ref{eq:zer*1}): any change in the radial position in the image plane, $\rho$, results in rapid variations of the phase of the integrand.

In our previous investigations (e.g. \cite{Turyshev-Toth:2021-caustics}), we have seen that, with the exception of the immediate vicinity of the caustic, integral (\ref{eq:zer*1}) may be treated using the method of stationary phase. Such a solution is very valuable, as it allows for simplified and more efficient numerical modeling of realistic imaging applications of the SGL of the extended Sun.  Of course, we may directly use the integral (\ref{eq:zer*1}) and evaluate it numerically. However, because of the high spatial frequency involved, $\alpha=k\sqrt{2r_g/r}=48.97~{\rm m}^{-1}$ from (\ref{eq:zerJ}), it may require an unrealistically high number of integration steps in the spatial domain, characterized by $\rho$, to obtain numerical convergence. Thus, an approximate solution of (\ref{eq:zer*1}) is of great practical value.

\subsection{Establishing the quartic equation}

To develop a solution of (\ref{eq:zer*1}), we use the method of stationary phase that we successfully used in past for similar purposes \cite{Turyshev-Toth:2017,Turyshev-Toth:2019,Turyshev-Toth:2020-extend,Turyshev-Toth:2021-multipoles,Turyshev-Toth:2021-caustics}.
To do that, we consider  the integral (\ref{eq:zer*1}) and recognize that its oscillatory behavior is driven by the phase, $\varphi$, that has the form:
{}
\begin{eqnarray}
\varphi(\rho, \phi)&=&-\Big(\alpha\rho\cos(\phi_\xi-\phi)+\beta_2\cos[2(\phi_\xi-\phi_s)]\Big).
\label{eq:zer*1*}
\end{eqnarray}
To form a stationary phase solution form (\ref{eq:zer*1}), we compute the derivatives $\varphi'=d\varphi/d\phi_\xi$ and  $\varphi''=d^2\varphi/d\phi_\xi^2$:
{}
\begin{eqnarray}
\varphi'=
\frac{d\varphi}{d\phi_\xi}&=&\alpha\rho\sin(\phi_\xi-\phi)+2\beta_2\sin[2(\phi_\xi-\phi_s)],
\label{eq:phi-der}\\
\varphi''=
\frac{d^2\varphi}{d\phi_\xi^2}&=&\alpha\rho\cos(\phi_\xi-\phi)+4\beta_2\cos[2(\phi_\xi-\phi_s)].
\label{eq:phi-der2}
\end{eqnarray}

The phase is stationary when its first derivative (\ref{eq:phi-der}) vanishes. By setting $\varphi'=d\varphi/d\phi_\xi=0$, we have the equation to find the values of $\phi_\xi$ for the stationary phase:
{}
\begin{eqnarray}
\alpha\rho\sin(\phi_\xi-\phi)+2\beta_2\sin[2(\phi_\xi-\phi_s)]=0.
\label{eq:zer0}
\end{eqnarray}
We define
{}
\begin{eqnarray}
\mu=2(\phi-\phi_s),
\label{eq:zer01_mu}
\end{eqnarray}
and transform (\ref{eq:zer0}) as
{}
\begin{eqnarray}
\alpha\rho\sin(\phi_\xi-\phi)+2\beta_2\Big(1-2\sin^2(\phi_\xi-\phi)\Big)\sin\mu+4\beta\sin(\phi_\xi-\phi)\cos(\phi_\xi-\phi)\cos\mu=0.
\label{eq:zer_01+}
\end{eqnarray}

Next, to solve (\ref{eq:zer_01+}),  we introduce the quantity  $x$:
{}
\begin{eqnarray}
\sin(\phi_\xi-\phi)=x, ~~~~ \cos(\phi_\xi-\phi)=\pm\sqrt{1-x^2},
\label{eq:zer01}
\end{eqnarray}
and rewrite (\ref{eq:zer_01+}) as follows:
{}
\begin{eqnarray}
\alpha \rho \,x+2\beta_2 \sin\mu\,(1-2x^2)\pm4\beta_2 \cos\mu\,x\sqrt{1-x^2}=0.
\label{eq:zer01=}
\end{eqnarray}

Expressions (\ref{eq:zer01}) have a clear meaning. With parameterizations (\ref{eq:note}), we see that (\ref{eq:zer01}) may be given in a form that explicitly shows their geometric nature, namely $\cos(\phi_\xi-\phi)=(\vec n _\xi\cdot \vec n)$ and $\sin(\phi_\xi-\phi)=|[\vec n\times[\vec n _\xi\times \vec n]]|$, which are the Cartesian projections of the observer's position vector with respect to the vector of the impact parameter. Thus, for a given observer's position, $\rho\vec n$, there will always be $b\vec n_\xi$ for which the  quantities (\ref{eq:zer01}) will take all the values $x \in[-1,1]$.

By isolating the term with the square root in (\ref{eq:zer01=}) and squaring the result, we rewrite this equation as
{}
\begin{eqnarray}
16\beta^2_2\, x^4-8\alpha\rho\,\beta_2 \sin\mu \, x^3+\Big((\alpha\rho)^2-16\beta^2_2\Big)x^2+4\alpha\rho\beta_2\sin\mu \, x+4\beta^2_2\sin^2\mu=0.
\label{eq:zer01=0}
\end{eqnarray}

If $\beta_2=0$, Eq.~(\ref{eq:zer01=0}) yields $x=0$ or $\sin(\phi_\xi-\phi)=0$, resulting in $\phi_\xi=\phi$, which is consistent with the fact that in a monopole gravitational field, light rays propagate in a plane (see discussion in Appendix~\ref{sec:mono-case}).

If $\beta_2\not=0$, we divide (\ref{eq:zer01=0}) by $16\beta^2_2$ and, defining
{}
\begin{eqnarray}
\eta=\frac{\alpha\rho}{4\beta_2},
\label{eq:zer03}
\end{eqnarray}
we transform equation (\ref{eq:zer01=0}) as
{}
\begin{eqnarray}
x^4-2\eta\sin\mu \, x^3+\big(\eta^2-1\big)x^2+\eta\sin\mu \, x+{\textstyle\frac{1}{4}}\sin^2\mu=0.
\label{eq:zer03*}
\end{eqnarray}

Eq.~(\ref{eq:zer03*}) is our main equation for analytic investigations of the optical properties of the quadrupole gravitational lens. This equation is quartic with respect to $x$  and may be solved analytically. Remember that $x$ must be real and be bounded as $|x|\leq1$ in order for $\phi_\xi-\phi$ to be real. We note that Eq.~(\ref{eq:zer03*}), which determines the stationary phase, does not depend on the parameters $\alpha$ and $\beta$ separately, but rather on their ratio $\eta$, given by (\ref{eq:zer03}), which is independent of wavelength. Thus, the basic geometry of the solution is fully determined by the gravitational field of the lens.

\subsection{Solving a generic quartic equation}
\label{sec:quart-gen}

Solving polynomial equations is a well-established area of algebra. Methods for solving linear and quadratic equations have been known since antiquity. Solutions to the cubic and quartic equations were first published in Gerolamo Cardano's {\em Ars Magna} \cite{Cardano:1545}. The quartic solution, in particular, was developed by Niccol\`o Tartaglia, who disclosed this solution to Cardano in confidence. According to the recollection of Cardano's assistant, Lodovico Ferrari, it was upon learning of another solution, by Scipione del Ferro, that Cardano considered himself no longer bound by oath to keep Tartaglia's solution in secret.

\begin{figure}
\includegraphics{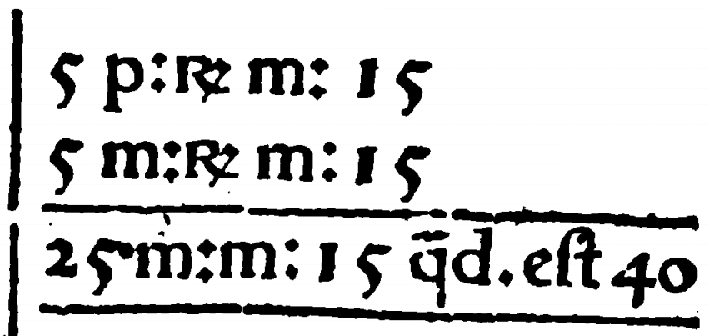}
\caption{\label{fig:cardano}Cardano's 1545 calculation of $(5 + \sqrt{-15})\times(5 - \sqrt{-15})$ as it appeared in {\em Ars Magna}.}
\end{figure}

The work of these exceptional mathematicians also led to the discovery of imaginary numbers. This was necessitated, in part, by the what is known as the {\em casus irreducibilis} in algebra: the fact that some real solutions to these equations can only be obtained by way of introducing complex numbers, even though in the end their imaginary parts cancel out, leaving only real terms. Perhaps troubled by this, Cardano wrote in {\em Ars Magna} the following: ``Dismissing mental tortures, and multiplying $5 + \sqrt{-15}$ by $5 - \sqrt{-15}$, we obtain $25 - (-15)$. Therefore the product is 40. .... and thus far does arithmetical subtlety go, of which this, the extreme, is, as I have said, so subtle that it is useless.''\footnote{``[...] \em{hucusque progreditur Arithmetica subtilitas, cuius hoc extremum ut dixi, adeo est subtile, ut fit inutile.}''} (Fig.~\ref{fig:cardano}.)

The quartic equation also turned out to be the highest degree algebraic equation that can be solved in the general case using only arithmetic and radicals. First proof that the general quintic equation has no algebraic solution was given in the form of the Abel--Ruffini theorem in 1824. The notes left by \'E. Galois led to an elegant group theoretical result showing that there for the quintic and above, there are always combinations of polynomial coefficients for which the symmetry properties of the roots of the equation are incompatible with the symmetry properties of numbers that can be expressed using only basic arithmetic and radicals \cite{Garling:1986,Stewart:2004}.

In the case of numerical algorithms on modern computers, the subtlety that Cardano found so vexing manifests itself in a different way. Imaginary parts cancel out additively, and the cancelation is often subject to rounding errors. Thus, a result that should be strictly real acquires a small but nonvanishing imaginary part. This can easily confuse numerical algorithms that, e.g., use the exponential representation of a complex number to calculate its roots. A small negative imaginary part, for instance, may be represented by the phase $2\pi-\epsilon$ (with $0<\epsilon\ll 1$) that will yield completely different values than a real-only algorithm for this quantity's square or cubic root.

Difficulties like this are among the many reasons why cubic and quartic equations are almost never solved algebraically when numerical solutions are readily available \cite{Szele:1975}. However, in our case, perhaps for the first time in our scientific careers, we found that studying the algebraic solution yields unprecedented insight. It is for this reason that we chose to proceed and work out the solution that is presented below.

In modern notation\footnote{To solve a quartic equation, we borrow the notation from \url{https://en.wikipedia.org/wiki/Quartic_function}.}, the four roots of the general quartic equation
{}
\begin{eqnarray}
a x^4+ b x^3+c x^2+dx+e=0,
\label{eq:zer04}
\end{eqnarray}
with $a\not=0$, have the following form:
{}
\begin{eqnarray}
x_{1,2}&=&-\frac{b}{4a}-S\pm\frac{1}{2}\sqrt{-4S^2-2p+\frac{q}{S}},\\
x_{3,4}&=&-\frac{b}{4a}+S\pm\frac{1}{2}\sqrt{-4S^2-2p-\frac{q}{S}},
\label{eq:zer04*}
\end{eqnarray}
where $p$ and $q$ are given as
{}
\begin{eqnarray}
p&=&\frac{8ac-3b^2}{8a^2},
\qquad q=\frac{b^3-4abc+8a^2d}{8a^3},
\label{eq:zer05*}
\end{eqnarray}
and $S$ and $Q$ have the form
{}
\begin{eqnarray}
S&=&\frac{1}{2}\Big[-\frac{2}{3}p+\frac{1}{3a}\Big(Q+\frac{\Delta_0}{Q}\Big)\Big]^{1/2},
\qquad
Q=\Big[\frac{1}{2}\Big(\Delta_1+\sqrt{\Delta^2_1-4\Delta^3_0}\Big)\Big]^{1/3},
\label{eq:zer06*}
\end{eqnarray}
where $\Delta_0$ and $\Delta_1$ are
{}
\begin{eqnarray}
\Delta_0&=&c^2-3bd+12ae,
\qquad
\Delta_1=2c^3-9bcd+27b^2e+27ad^2-72ace.
\label{eq:zer07*}
\end{eqnarray}

We will use this solution (\ref{eq:zer04*}) with (\ref{eq:zer05*})--(\ref{eq:zer07*}) to find the roots of the quartic equation (\ref{eq:zer03*}).

The quantity
\begin{align}
\Delta=-\tfrac{1}{27}(\Delta^2_1-4\Delta^3_0)\label{eq:disc}
\end{align}
that appears in (\ref{eq:zer06*}) is recognized as the discriminant of the quartic equation. When $\Delta>0$ either all four roots of the quartic equation are real or all four roots are complex; when $\Delta\le 0$, there will be typically two real roots (except for special cases, when some roots coincide).

\subsection{The quartic equation for the quadrupole gravitational lens}
\label{sec:quartic}

\begin{figure}
\begin{tabular}{p{0.4\textwidth} p{0.5\textwidth}}
\vspace{0in}\includegraphics{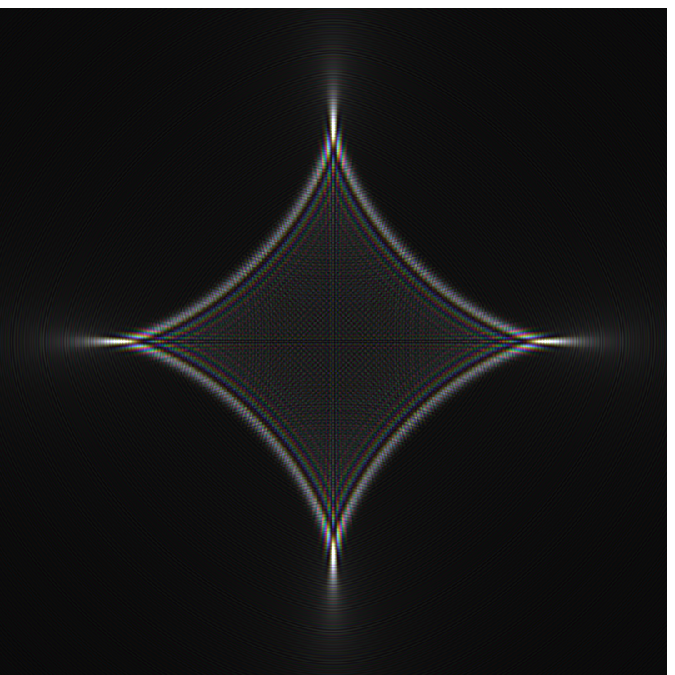} &
\vspace{-0.5in}\includegraphics{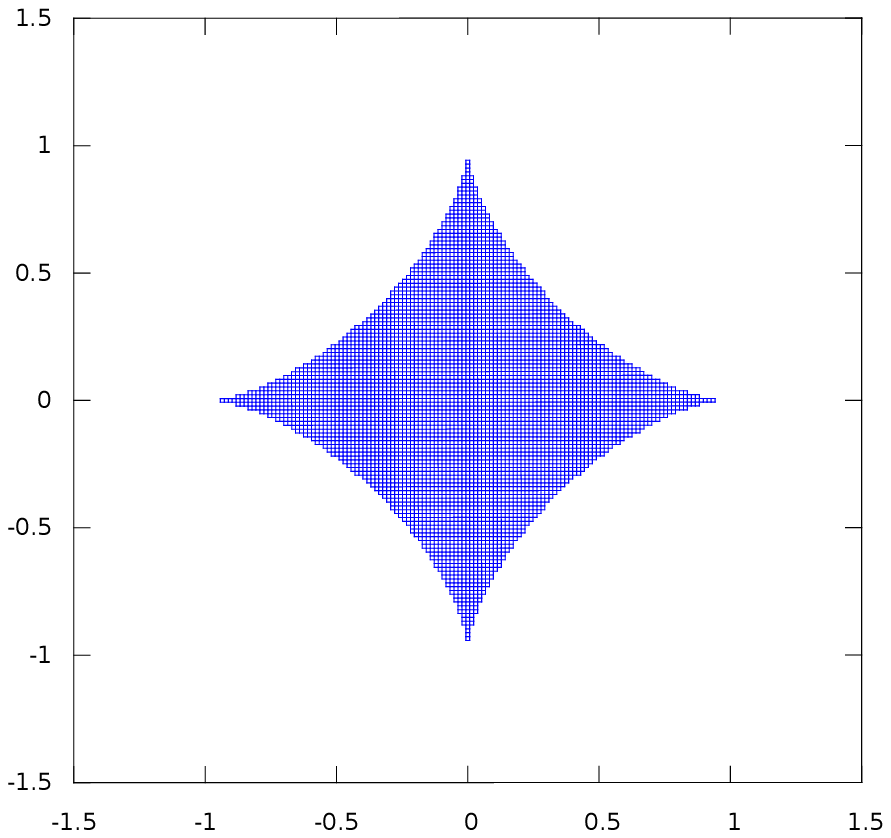}
\end{tabular}
\vskip -2em
\caption{\label{fig:disc}Left: a typical astroid caustic, produced by the quadrupole zonal harmonic, $J_2$, of a gravitational lens (from \cite{Turyshev-Toth:2021-multipoles}). Right: the region in the image plane, characterized by dimensionless coordinates
$(\eta\cos\tfrac{1}{2}\mu,\eta\sin\frac{1}{2}\mu)$, where the quartic discriminant $\Delta$ from (\ref{eq:disc}), is given by (\ref{eq:discq}), for the parameters in the equation (\ref{eq:zer03*})  is positive.}
\end{figure}

Investigating (\ref{eq:zer03*}), we first identify the relevant coefficients present in the generic quartic equation (\ref{eq:zer04}), namely
{}
\begin{eqnarray}
a=1, \qquad b=-2\eta\sin\mu, \qquad c=\eta^2-1, \qquad
d=  \eta\sin\mu,\qquad
e={\textstyle\frac{1}{4}}\sin^2\mu,
\label{eq:gen-q}
\end{eqnarray}
where $\mu$ and $\eta$ are given by (\ref{eq:zer01_mu}) and (\ref{eq:zer03}), correspondingly.  Furthermore, we note that these two quantities, $\frac{1}{2}\mu=\phi-\phi_s$ and $\eta=(\alpha/4\beta_2)\rho$,
may serve as a system of polar coordinates in the image plane. Therefore, we expect the solutions of Eq.~(\ref{eq:zer03*}) to be representable using image plane coordinates.

To study the possible roots of (\ref{eq:zer03*}), we first turn to the discriminant (\ref{eq:disc}), which, using (\ref{eq:gen-q}), is computed to be:
\begin{align}
\Delta (\eta,\mu)=-\Big((\eta^2-1)^3+\tfrac{27}{4}\eta^4\sin^2\mu\Big)\sin^2\mu.\label{eq:discq}
\end{align}
Figure~\ref{fig:disc} shows the region where $\Delta>0$, i.e., where Eq.~(\ref{eq:zer03*}) is expected to have either four real or four complex roots. Remarkably, we immediately recognize the shape of the astroid caustic associated with the quadrupole gravitational lens, familiar from our earlier studies \cite{Turyshev-Toth:2021-multipoles,Turyshev-Toth:2021-caustics}. This can be confirmed by a small amount of algebra, verifying that the boundary $\Delta=0$ is solved by values of $\eta$ and $\mu=2(\phi-\phi_s)$ that satisfy the defining equation of the astroid in polar coordinates, $\eta=1/(\cos^{2/3}\tfrac{1}{2}\mu+\sin^{2/3}\tfrac{1}{2}\mu)^{3/2}$. As the four solutions of (\ref{eq:zer03*}) for $\eta=0$, $x^4-x^2+\frac{1}{4}\sin^2\mu=0$, given by $x_{1..4}=\pm\sqrt{(1\pm\cos{\mu})/2}$, are obviously real, we conclude that this astroid region is characterized by four real roots.

To obtain the actual values of these roots, and to investigate the behavior of the roots outside the caustic boundary, we apply the full solution to the generic quartic equation, summarized above in Sec.~\ref{sec:quart-gen}, and present the four solutions,   $x_n(\eta,\mu), n\in[1,4]$ as
{}
\begin{eqnarray}
x_{1,2}(\eta,\mu)&=&{\textstyle\frac{1}{2}}\eta\sin\mu-S\pm{\textstyle\frac{1}{2}}\Big(-4S^2-2(\eta^2-1)+3\eta^2\sin^2\mu+\frac{\eta^3\cos^2\mu\sin\mu}{S}\Big)^\frac{1}{2},
\label{eq:sol01}\\
x_{3,4}(\eta,\mu)&=&{\textstyle\frac{1}{2}}\eta\sin\mu+S\pm{\textstyle\frac{1}{2}}\Big(-4S^2-2(\eta^2-1)+3\eta^2\sin^2\mu-\frac{\eta^3\cos^2\mu\sin\mu}{S}\Big)^\frac{1}{2},
\label{eq:sol01b}
\end{eqnarray}
where the quantities $S=S(\eta,\mu)$ and $Q=Q(\eta,\mu)$ are given by
{}
\begin{eqnarray}
S&=&{\textstyle\frac{1}{2}}\Big\{-{\textstyle\frac{2}{3}}(\eta^2-1)+\eta^2\sin^2\mu+{\textstyle\frac{1}{3}}\Big(Q+\frac{(\eta^2-1)^2+3(2\eta^2+1)\sin^2\mu}{Q}\Big)\Big\}^\frac{1}{2},\\
Q&=&\Big\{(\eta^2-1)^3+9(\eta^2-1)^2\sin^2\mu+{\textstyle\frac{27}{2}}\eta^2\sin^2\mu\big(1+\sin^2\mu\big)+
\nonumber \\
&&\hskip 43pt~+
\cos^2\mu\Big[\Big(27(\eta^2-1)^3+\big({\textstyle\frac{27}{2}}\big)^2\eta^4\sin^2\mu\Big)\sin^2\mu\Big]^\frac{1}{2}\Big\}^\frac{1}{3}.~~~~~
\label{eq:sol02}
\end{eqnarray}

We will use the solutions (\ref{eq:sol01})--(\ref{eq:sol02}) to evaluate the integral (\ref{eq:zer*1}) in Section~\ref{sec:quartic}.
However, before doing that, we note that what we are actually solving for is $\phi$ through Eq.~(\ref{eq:zer01}), and that due to the nature of the trigonometric function, an ambiguity remains as $\sin(\phi_\xi-\phi)=\sin\big(\pi-(\phi_\xi-\phi)\big)$.

This ambiguity could be resolved if we knew the value of $\cos(\phi_\xi-\phi)=\big(1-\sin^2(\phi_\xi-\phi)\big)^\frac{1}{2}$. To resolve this, we present an alternate form of the quartic equation, written in terms of $y=\sqrt{1-x^2}=\cos(\phi_\xi-\phi)$ (see also Appendix~\ref{sec:cos}):
{}
\begin{eqnarray}
y^4+2\eta \cos\mu \, y^3+(\eta^2-1)y^2-2\eta\cos\mu \, y+{\textstyle\frac{1}{4}}\sin^2\mu-\eta^2=0.
\label{eq:zer0as0copy}
\end{eqnarray}
Visual inspection of this equation tells us that $y=0$ if and only if $\tfrac{1}{4}\sin^2\mu-\eta^2=0$. The sign of this expression, therefore, can determine the sign of $y=\cos(\phi_\xi-\phi)$.

Relying on this knowledge, and sampling the solution space for $x$ and $y$ in the $(\eta,\mu)$ plane allows us to determine the respective signs of $x_i$ and $y_i=\sqrt{1-x_i^2}$:

\begin{minipage}{0.34\textwidth}
\begin{align}
\underline{\sin(\phi_{\xi}-\phi)}&=x_1(\eta, \mu), \nonumber\\
\mu\in\,]0,{\textstyle\frac{1}{2}}\pi[:~\cos(\phi_{\xi}-\phi)&=\sqrt{1-x_1^2}, \nonumber\\
\mu\in\,]{\textstyle\frac{1}{2}}\pi,\pi[:~\cos(\phi_{\xi}-\phi)&=-\sqrt{1-x_1^2},
\nonumber\\
\mu\in\,]\pi,{\textstyle\frac{3}{2}}\pi[:~\cos(\phi_{\xi}-\phi)&=\sqrt{1-x_1^2},
\nonumber\\
\mu\in\,]{\textstyle\frac{3}{2}}\pi,2\pi[:~\cos(\phi_{\xi}-\phi)&=-\sqrt{1-x_1^2}.
\label{eq:sol1}\nonumber\\
~\nonumber\\
~\nonumber\\
~
\end{align}
\end{minipage}
\begin{minipage}{0.64\textwidth}
\begin{align}
\underline{\sin(\phi_{\xi}-\phi)}&=x_2(\eta, \mu), \nonumber\\
\mu\in\,]0,{\textstyle\frac{1}{2}}\pi[:~\cos(\phi_{\xi}-\phi)&=-\sqrt{1-x_2^2}, \nonumber\\
\mu\in\,]{\textstyle\frac{1}{2}}\pi,\pi[:~\cos(\phi_{\xi}-\phi)&=\sqrt{1-x_2^2},
\nonumber\\
\mu\in\,]\pi,{\textstyle\frac{3}{2}}\pi[~\wedge~\tfrac{1}{4}\sin^2\mu-\eta^2\ge 0:~\cos(\phi_{\xi}-\phi)&=-\sqrt{1-x_2^2},
\nonumber\\
\mu\in\,]\pi,{\textstyle\frac{3}{2}}\pi[~\wedge~\tfrac{1}{4}\sin^2\mu-\eta^2<0:~\cos(\phi_{\xi}-\phi)&=\sqrt{1-x_2^2},
\nonumber\\
\mu\in\,]{\textstyle\frac{3}{2}}\pi,2\pi[~\wedge~\tfrac{1}{4}\sin^2\mu-\eta^2\ge 0:~\cos(\phi_{\xi}-\phi)&=\sqrt{1-x_2^2},
\nonumber\\
\mu\in\,]{\textstyle\frac{3}{2}}\pi,2\pi[~\wedge~\tfrac{1}{4}\sin^2\mu-\eta^2<0:~\cos(\phi_{\xi}-\phi)&=-\sqrt{1-x_2^2}.
\label{eq:sol2}
\end{align}
\end{minipage}\\
~\\
\begin{minipage}{0.64\textwidth}
\begin{align}
\underline{\sin(\phi_{\xi}-\phi)}&=x_3(\eta, \mu), \nonumber\\
\mu\in\,]0,{\textstyle\frac{1}{2}}\pi[~\wedge~\tfrac{1}{4}\sin^2\mu-\eta^2\ge 0:~\cos(\phi_{\xi}-\phi)&=\sqrt{1-x_3^2},
\nonumber\\
\mu\in\,]0,{\textstyle\frac{1}{2}}\pi[~\wedge~\tfrac{1}{4}\sin^2\mu-\eta^2<0:~\cos(\phi_{\xi}-\phi)&=-\sqrt{1-x_3^2}, \nonumber\\
\mu\in\,]{\textstyle\frac{1}{2}}\pi,\pi[~\wedge~\tfrac{1}{4}\sin^2\mu-\eta^2\ge 0:~\cos(\phi_{\xi}-\phi)&=-\sqrt{1-x_3^2},
\nonumber\\
\mu\in\,]{\textstyle\frac{1}{2}}\pi,\pi[~\wedge~\tfrac{1}{4}\sin^2\mu-\eta^2<0:~\cos(\phi_{\xi}-\phi)&=\sqrt{1-x_3^2},
\nonumber\\
\mu\in\,]\pi,{\textstyle\frac{3}{2}}\pi[:~\cos(\phi_{\xi}-\phi)&=\sqrt{1-x_3^2},
\nonumber\\
\mu\in\,]{\textstyle\frac{3}{2}}\pi,2\pi[:~\cos(\phi_{\xi}-\phi)&=-\sqrt{1-x_3^2},
\label{eq:sol3}
\end{align}
\end{minipage}
\begin{minipage}{0.36\textwidth}
\begin{align}
\underline{\sin(\phi_{\xi}-\phi)}&=x_4(\eta, \mu), \nonumber\\
\mu\in\,]0,{\textstyle\frac{1}{2}}\pi[:~\cos(\phi_{\xi}-\phi)&=-\sqrt{1-x_4^2}, \nonumber\\
\mu\in\,]{\textstyle\frac{1}{2}}\pi,\pi[:~\cos(\phi_{\xi}-\phi)&=\sqrt{1-x_4^2},
\nonumber\\
\mu\in\,]\pi,{\textstyle\frac{3}{2}}\pi[:~\cos(\phi_{\xi}-\phi)&=-\sqrt{1-x_4^2},
\nonumber\\
\mu\in\,]{\textstyle\frac{3}{2}}\pi,2\pi[:~\cos(\phi_{\xi}-\phi)&=\sqrt{1-x_4^2}.
\label{eq:sol4}\nonumber\\
~\nonumber\\
~\nonumber\\[2pt]
\end{align}%
\end{minipage}

~

As a result, we were able to identify all the pairs of the solutions of the quartic equation  (\ref{eq:zer03*}). At this point, we can use these results in the evaluation of the integral (\ref{eq:zer*1}).

\begin{figure}
\vskip -0.75em
\begin{tabular}{p{0.5\textwidth} p{0.5\textwidth}}
\hskip -2em
\includegraphics{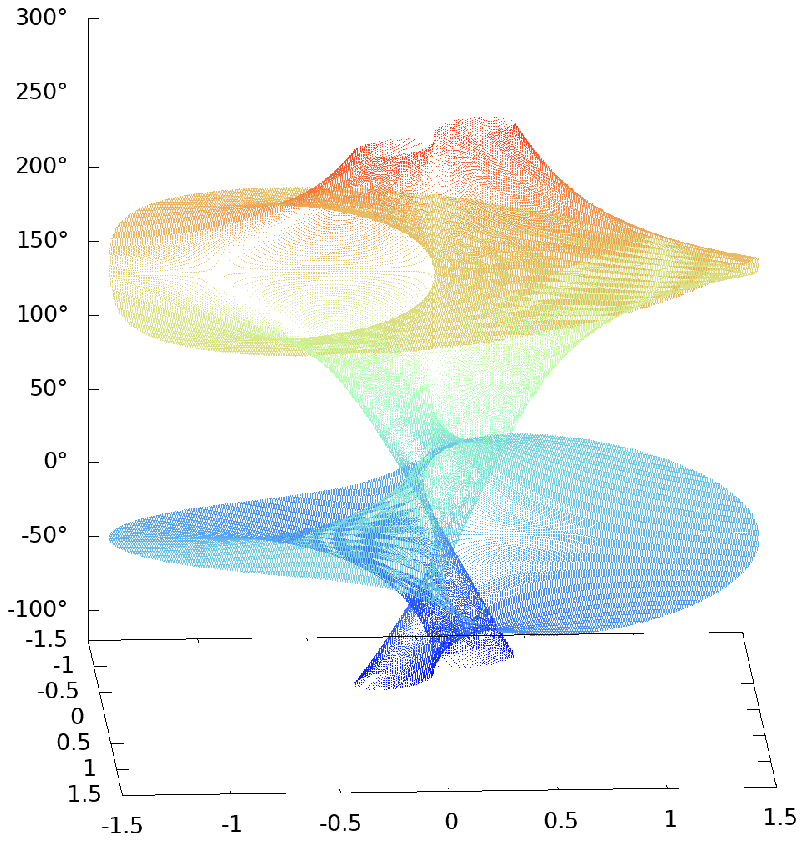} &
\hskip -3em
\includegraphics{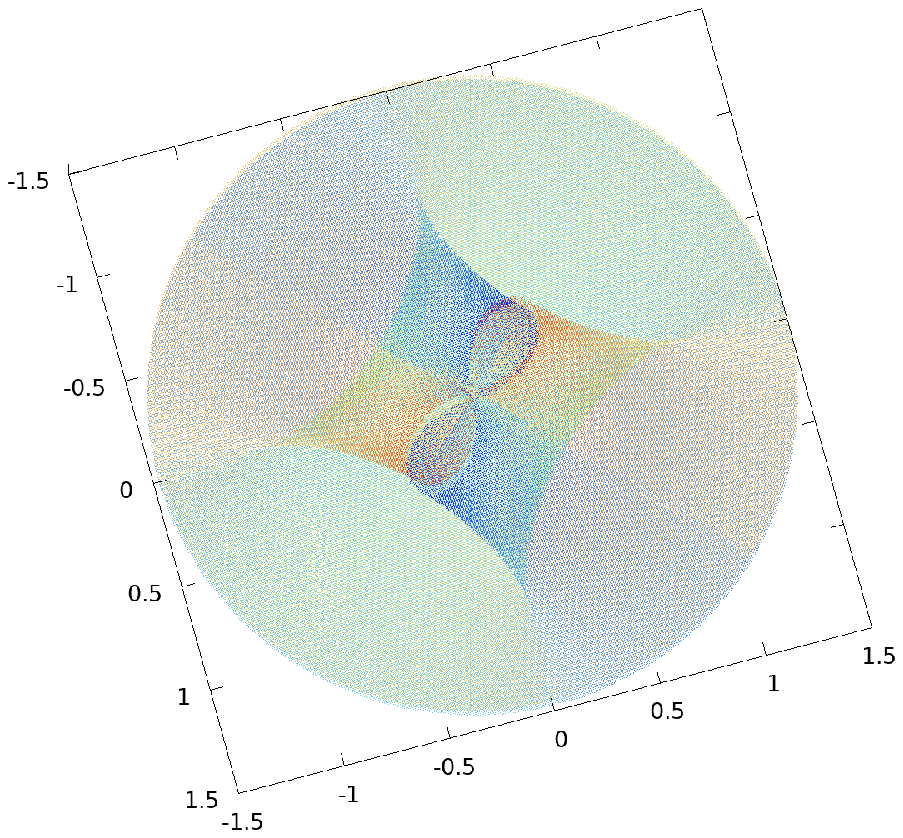}
\end{tabular}
\vskip -1.25em
\caption{\label{fig:phi}Plotting the real roots of the quartic equation (\ref{eq:zer03*}), expressed as the angle $(\phi_\xi-\phi)$ (see Eq.~(\ref{eq:zer01})) as  functions of the parameters $\eta$ and $\mu$, for $\eta\le 1.5$. Left: a perspective view of the four solutions. The horizontal plane is characterized by dimensionless coordinates $(\eta\cos\tfrac{1}{2}\mu,\eta\sin\tfrac{1}{2}\mu)$; the vertical axis represents the values of the roots, $(\phi_\xi-\phi)$, in degrees. Right: viewing the solutions from a direction characterized by $\vec{k}$, i.e., perpendicular to the $(\eta,\mu)$ image plane. The astroid region in which four real roots exist, as predicted by the behavior of the discriminant $\Delta$ given by (\ref{eq:discq}), is clearly visible from this viewing direction. The image is tilted to reduce the occurrence of display artifacts that may appear when viewing. Color is used only for better visualization. For values of $1.5<\eta\to\infty$, the result quickly settles on two angles, $180^\circ$ apart. For a video animation, see \protect\url{https://vttoth.com/CMS/physics-notes/358}.}
\end{figure}

What does this solution space look like? Having obtained both $\sin(\phi_\xi-\phi)$ and $\cos(\phi_\xi-\phi)$, we are now in a position to plot the quantity $(\phi_\xi-\phi)$ in the image plane, characterized by $\eta=(\alpha/4\beta_2)\rho$ as the dimensionless radial coordinate, and $\phi$ as the angular coordinate. The result, shown in Fig.~\ref{fig:phi} (left), is instructive.

We see that the central region of the three-dimensional structure represented by the four roots is rather complex. To make more sense of this region, we can also look at it ``from the above'', i.e., from a viewing direction that is perpendicular to the image plane, as shown in Fig.~\ref{fig:phi} (right). This confirms our prediction that we made on the basis of inspecting the quartic discriminant $\Delta$: the cross-sectional shape of this region is that of the astroid that characterizes the quadrupole gravitational lens.

Outside this astroid region, in any direction, only two real roots remain. Looking at Fig.~\ref{fig:phi} (left), we may already surmise a pattern: outside the astroid, the two remaining real solutions represent two angles that are $180^\circ$ apart.

\begin{figure}
\rotatebox{90}{\hskip 0.75in$\phi_\xi-\phi$}\includegraphics{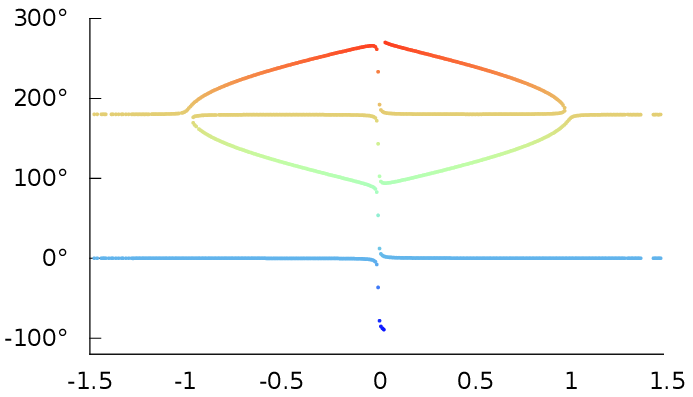}~\rotatebox{90}{\hskip 0.75in$\phi_\xi-\phi$}\includegraphics{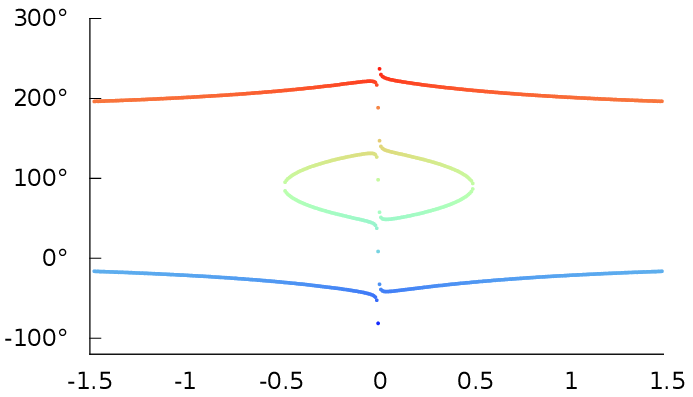}\\
\vskip -1.5em
$\eta$\hskip 0.44\textwidth $\eta$\\
\caption{\label{fig:crossdir}Cross-sectional representation of the quartic equation solution space show in Fig.~\ref{fig:phi}, as viewed in the direction of the astroid cusp (left), characterized by $\mu=0$; and in the direction of the astroid fold (right), characterized by $\tfrac{1}{2}\mu=\pi/4$. Horizontal axis is $\eta$. The vertical axis represents the roots of the equation, expressed in degrees as $(\phi_\xi-\phi)$  corresponding to Eq.~(\ref{eq:zer01}). We can see that depending on the value of $\eta$ there are either four real roots (in the interior of the caustic region) or only two (outside the caustic region).}
\end{figure}

To investigate limiting cases of the solutions (\ref{eq:sol01})--(\ref{eq:sol02}) we can look at plots that represent cross-sectional views of the solution space depicted in Fig.~\ref{fig:phi}. In Fig.~\ref{fig:crossdir} (left) we see the solutions both inside ($|\eta|\le 1$) and outside the caustic region. The cusp position is given by the values of $\eta=\pm1$, see \cite{Turyshev-Toth:2021-caustics}. Inside, we have four roots; outside, we have two roots, which, for large $\eta$, appear to converge on values that are $180^\circ$ apart.

The cross-sectional view in the direction of the astroid fold, $\tfrac{1}{2}\mu=\pi/4$, shown in Fig.~\ref{fig:crossdir} (right), is equally revealing. (The fold position is given by $\eta=\pm \tfrac{1}{2}$, see \cite{Turyshev-Toth:2021-caustics}.) In this direction, the caustic boundary is at $|\eta|=\tfrac{1}{2}$ from the origin. Indeed, we see that of the four roots of the quartic equation, two vanish exactly on the caustic boundary; outside, far from the origin, the remaining two roots rapidly settle down on two values that are again approximately $180^\circ$ apart.

We can confirm this behavior outside the caustic boundary by inspecting the roots themselves.

In the limit of large $\eta$, the quantity $Q$ behaves as $\lim\limits_{\eta\to\infty}Q=\eta^2$ and, therefore, $S=\tfrac{1}{2}\eta|\sin\mu|$. Taking the limits of the roots, we obtain:
\begin{align}
\begin{cases}
\lim\limits_{\eta\to\infty}x_{1,2}=0, & \text{when } \sin\mu \ge 0,\\
\lim\limits_{\eta\to\infty}x_{3,4}=0, & \text{when } \sin\mu \le 0,
\end{cases}\label{eq:tworootcases}
\end{align}
with the remaining roots complex. Furthermore, we find that the two real roots have opposite signs in these limits. Recalling that $x=\sin(\phi_\xi-\phi)$ and inspecting the corresponding cosine using (\ref{eq:sol1})--(\ref{eq:sol4}), we can see that indeed, the two remaining roots correspond to two angles that are $180^\circ$ apart. Looking at (\ref{eq:zer0as0copy}) we can immediately see that in the large $\eta$ limit, it is solved by $y=\cos(\phi_\xi-\phi)=\pm 1$. The physical meaning of $(\phi_\xi-\phi)$ will become clear shortly; for now, we note that the limit $\sin(\phi_\xi-\phi)\to 0$ corresponds to the monopole PSF, as discussed in Appendix~\ref{sec:mono-case}.

\subsection{Assembling the solution for the complex amplitude $B(\rho, \phi)$}

Before we begin assembling the solution for $B(\vec x)$, we recall that quantities $\alpha$, $\beta_2$, $\eta$, $\mu$ and $x$ are given by (\ref{eq:zerJ}), (\ref{eq:zer03}), (\ref{eq:zer01_mu}), and (\ref{eq:zer01}), correspondingly. They relate the solution (\ref{eq:sol1})--(\ref{eq:sol4}), to the quantity of interest, namely the specific eikonal phase $\phi_\xi$. With the four solutions for $x_n(\eta,\phi), \, n\in[1,4]$, we may write the formal solution for the integral (\ref{eq:zer*1}) in the following form:
{}
\begin{eqnarray}
B_2(\rho, \phi)&=&
\frac{1}{\sqrt{2\pi}}\sum_{n=1}^4
\frac{1}{\sqrt{|\varphi''_n|}}
e^{i\big(\varphi_n+{\rm sign}[\varphi''_n] {\textstyle\frac{\pi}{4}}\big)},~~~
\label{eq:sol*1}
\end{eqnarray}
where the phase of the integral (\ref{eq:zer*1}), computed for each of the solutions (\ref{eq:sol1})--(\ref{eq:sol4}), $\varphi_n(\eta,\mu)$, and its second derivative also evaluated for each of these four solutions, $\varphi''_n(\eta,\mu)$, are computed from (\ref{eq:phi-der}) and (\ref{eq:phi-der2}), by substituting solutions for $x_n(\eta,\mu), \, n\in[1,4]$ from (\ref{eq:sol01})--(\ref{eq:sol02}). As for the signs in the phase of this expression, we take  the minus sign is the case when $\varphi''_n(\eta,\mu)>0$ and the plus sign is for $\varphi''_n(\eta,\mu)<0$.  Clearly, as the signs of $\varphi''_n(\eta,\mu)$ are changing as a function of a particular point on the image plane, these expressions will have to change signs accordingly.  Thus,  ${\rm sign}[\varphi''_n(\eta,\mu)]$ is also a function of $\eta$ and $\mu$. Such a smooth dependence complicates analytical formulation of the integral on the entire $(\eta,\mu)$-plane, although it can be done piecewise, or performed algorithmically for numerical plotting.

We evaluate the phase of the integral $\varphi_n(\eta,\mu)$ for each of the solutions (\ref{eq:sol1})--(\ref{eq:sol4}),  and,  using (\ref{eq:zer*1*}) and (\ref{eq:zer01}), we present it as
{}
\begin{eqnarray}
\varphi_n(\eta,\mu)&=&-\Big(\alpha\rho\cos(\phi_\xi-\phi)+\beta_2\cos[2(\phi_\xi-\phi_s)]\Big)=\nonumber\\
&=&4\beta_2\Big[\mp\Big(\eta-{\textstyle\frac{1}{2}} x_n(\eta,\mu) \sin\mu\Big)\sqrt{1-x_n^2(\eta,\mu)} -{\textstyle\frac{1}{4}}\Big(1-2x_n^2(\eta,\mu)\Big)\cos\mu\Big].
\label{eq:zer*1*n}
\end{eqnarray}
Similarly, for $\varphi''_n(\eta,\mu)$, with (\ref{eq:phi-der2}) and (\ref{eq:zer01}), we have
{}
\begin{eqnarray}
\varphi''_n(\eta,\mu)&=&\alpha\rho\cos(\phi_\xi-\phi)+4\beta_2\cos[2(\phi_\xi-\phi_s)]=\nonumber\\
&=&4\beta_2\Big[\pm\Big(\eta-2x_n(\eta,\mu)\sin\mu\Big)\sqrt{1-x_n^2(\eta,\mu)}+\Big(1-2x_n^2(\eta,\mu)\Big)\cos\mu\Big],
\label{eq:phi-der2*}
\end{eqnarray}
where the choice of the signs is set by the solution (\ref{eq:sol1})--(\ref{eq:sol4}).

As a result, we may now compute the light amplification factor of the lens in the presence of the quadrupole. For this, we substitute (\ref{eq:sol*1}) in (\ref{eq:zer*1}) and then into (\ref{eq:DB-sol-rho}). We then compute the Poynting vector and normalize it to get  the amplification in the presence of $J_2$:
{}
\begin{eqnarray}
\mu_{J_2}&=&2\pi k r_gB^2_2(\rho,\phi) \equiv 2\pi k r_g\, {\rm PSF},
\label{eq:mu}
\end{eqnarray}
where ${\rm PSF}=B_2^2(\rho,\phi)\equiv B_2(\rho,\phi)B_2^*(\rho,\phi)$
(with $B^*(\rho,\phi)$ is a complex conjugate of $B(\rho,\phi)$) is the PSF of the SGL that in the case when the quadrupole is present has the form:
{}
\begin{eqnarray}
{\rm PSF}&=&\frac{1}{2\pi}\Big\{\Big(\sum_{n}\frac{\cos[\varphi_n+{\rm sign}[\varphi''_n] {\textstyle\frac{\pi}{4}}] }{\sqrt{|\varphi''_n|}}\Big)^2+\Big(\sum_{n}\frac{\sin[\varphi_n+{\rm sign}[\varphi''_n]
{\textstyle\frac{\pi}{4}}]}{\sqrt{|\varphi''_n|}}\Big)^2\Big\},
\label{eq:PSF}
\end{eqnarray}
where the summation is over the real roots of the quartic equation.

\begin{figure}
\includegraphics{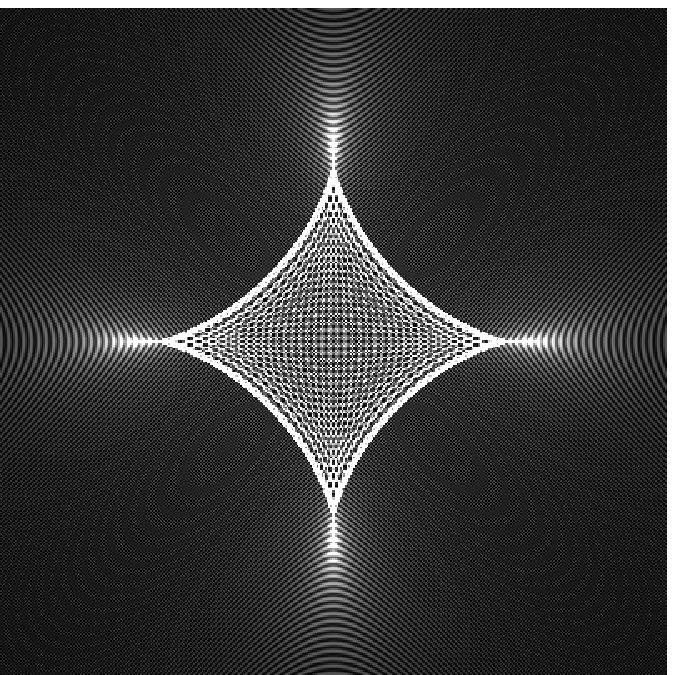}\hskip 0.5in\includegraphics{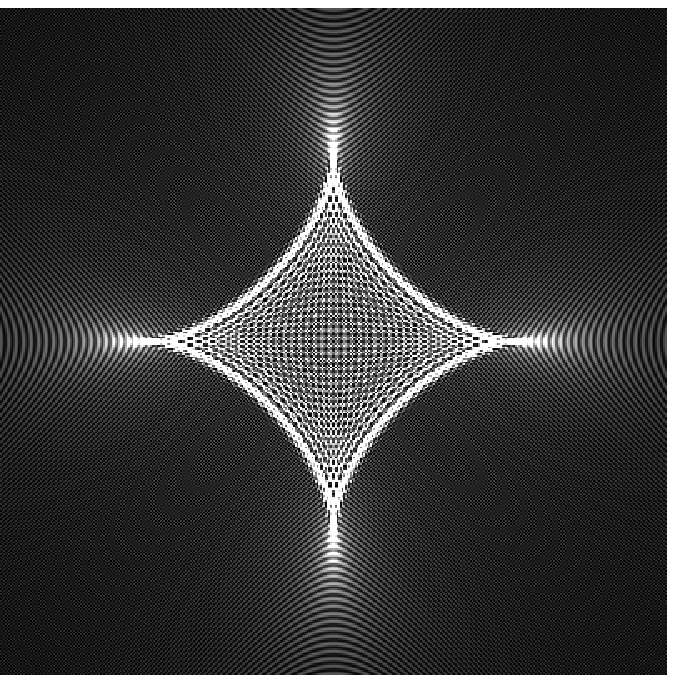}
\caption{\label{fig:astroid}The astroid caustic as calculated using the quartic solution (left) and numerical evaluation of (\ref{eq:B2}) (right).}
~\par
\includegraphics{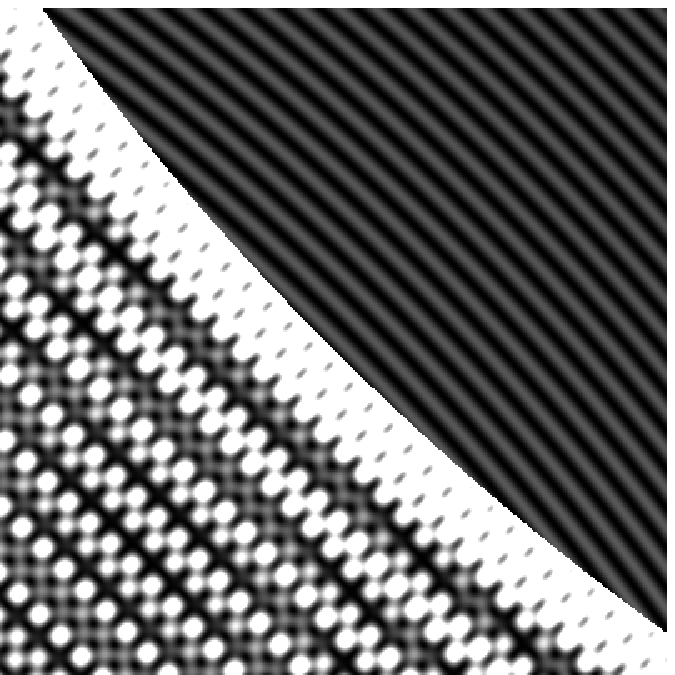}\hskip 0.5in\includegraphics{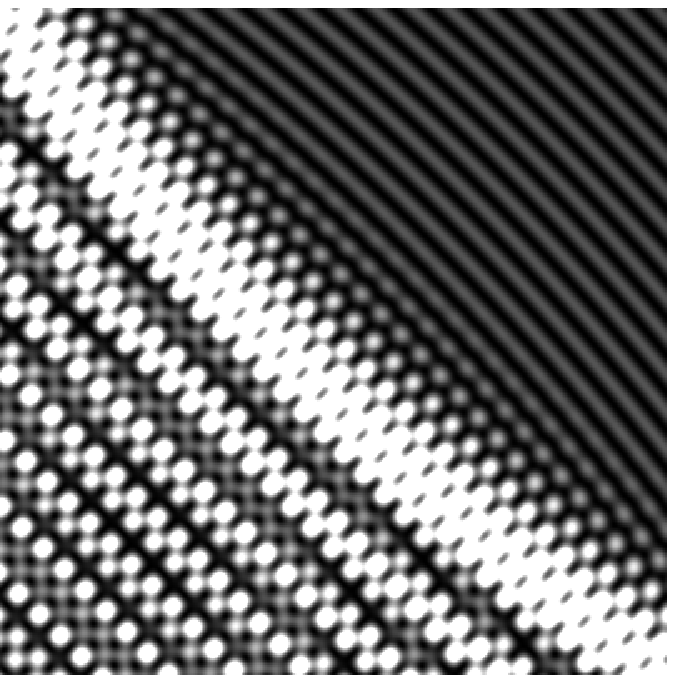}
\caption{\label{fig:astrdet}Detail of the astroid caustic from Fig.~\ref{fig:astroid} near the fold, as calculated using the quartic solution (left) and numerical evaluation of (\ref{eq:B2}) (right).}
\end{figure}

Figure~\ref{fig:astroid} (left) shows the resulting PSF, plotted using parameters characteristic for the Sun, at a focal plane distance of 650~AU, in a $20\times 20$ meter region, for $\sin^2\beta_s=0.02$. (On a practical note, it is worth mentioning that as the real roots switch between $x_{1,2}$ and $x_{3,4}$ on boundaries characterized by $\tfrac{1}{2}\mu=\tfrac{1}{2}n\pi$ in accordance with (\ref{eq:tworootcases}), computing (\ref{eq:PSF}) precisely on these boundaries can introduce numerical artifacts as quantities of opposite sign, rounded differently by machine algorithms, fail to cancel exactly. These situations can be avoided by shifting the origin of the coordinate system used for computation by an imperceptible amount, a small fraction of the computed pixel size.)

As noted above, the caustic boundary is marked by the transition from four to two roots in the quartic solution. This suggests a sudden, sharp boundary. This is indeed what we observe in Fig.~\ref{fig:astroid} (left). This can be compared against direct numerical evaluation of Eq.~(\ref{eq:B2}) as shown in Fig.~\ref{fig:astroid} (right). The absence of a sharp boundary is evident.

To see it in more detail, consider Figure~\ref{fig:astrdet}. This figure shows an enlarged, $2\times 2$ meter region of the astroid caustic in the direction of the fold at $\tfrac{1}{2}\mu=\tfrac{1}{4}\pi$. The quartic solution is characterized by a very sharp boundary, an instantaneous drop in light intensity, outside of which we see the emergence of the concentric pattern that is characteristic of a monopole lens. Numerical integration, on the other hand, demonstrates that the drop in intensity is more gradual, with a transitional region that extends to both sides of the caustic boundary. We also see that outside this very narrow region alongside the caustic boundary, the quartic solution very accurately reproduces the solution obtained through numerical integration, at a fraction of computational effort.

\section{Imaging point sources with a quadrupole lens}
\label{sec:quart-inaging}

At this point we are ready to embark on the most interesting part of the our investigation -  the formation and the evolution of the Einstein cross formed by the PSF developed from the previous section, e.g., (\ref{eq:PSF}).  Therefore, to reach the objectives of our investigation of the analytic solution of the quadrupole gravitational lens, (\ref{eq:sol1})--(\ref{eq:sol4}), we now focus our attention on developing a view that is seen through a thin lens optical telescope.

\subsection{PSF at different telescope positions}

Following \cite{Turyshev-Toth:2020-image,Turyshev-Toth:2021-imaging}, to describe imaging with the SGL of the extended Sun, we consider the SGL's PSF that is given in the form (\ref{eq:zer*1}). To convolve this PSF with that of a thin lens that represents an aperture of a telescope, we first need to establish an appropriate form of the PSF for extended sources. Examining the integral given by (\ref{eq:zer*1}), we see that it contains the following expression $\rho\cos(\phi_\xi-\phi)$ that may be transformed as
{}
\begin{eqnarray}
\rho\cos(\phi_\xi-\phi)=({\vec n}_\xi\cdot{\vec x}),~~~{\rm where}~~~ {\vec x}=\rho(\cos\phi, \sin\phi, 0).
\label{eq:rn}
\end{eqnarray}

Defining $\vec x$ to be the position of the telescope in the image plane and $\vec x'$ to be the any point in the same plane, we present $\vec x$ as
{}
\begin{eqnarray}
\vec x \qquad \Rightarrow \qquad {\vec x}+\vec x',~~~{\rm where}~~~ {\vec
x}'=\rho'(\cos\phi', \sin\phi', 0).
\label{eq:rn0}
\end{eqnarray}
This allows us to have the following transformation:
{}
\begin{eqnarray}
\rho\cos(\phi_\xi-\phi)=\big({\vec n}_\xi\cdot{\vec x}\big)
\qquad\Rightarrow \qquad
({\vec n}_\xi\cdot\big({\vec x}+{\vec x}')\big)=\rho\cos(\phi_\xi-\phi)+\rho'\cos(\phi_\xi-\phi').
\label{eq:rn2}
\end{eqnarray}

This leads to the following expression for the amplitude of the EM field:
{}
\begin{eqnarray}
{B}({\vec x}, {\vec x}')&=&\frac{1}{{2\pi}}\int_0^{2\pi} d\phi_\xi \exp\Big[-i\Big(\alpha\big(\rho\cos(\phi_\xi-\phi)+\rho'\cos(\phi_\xi-\phi')\big)+\beta_2\cos[2(\phi_\xi-\phi_s)]\Big)\Big].
\label{eq:zer*1_amp}
\end{eqnarray}

We rcall that the geometry is described by several parameters, such as ${\vec x}$ being the current position of an optical telescope in the SGL's image plane, ${\vec x}'$, being any point on the same plane, and ${\vec x}_i$, being a point on the focal plane of the optical telescope. These positions are given as
{}
\begin{eqnarray}
\{{\vec x}\}&\equiv& (x,y,0)=
\rho\big(\cos\phi,\sin\phi,0\big)=\rho{\vec n}, \label{eq:coord'}\\
\{{\vec x}'\}&\equiv& (x',y',0)=
\rho'\big(\cos\phi',\sin\phi',0\big)=\rho'\,{\vec n}',
\label{eq:x-im} \\
 \{{\vec x}_i\}&\equiv& (x_i,y_i,0)=\rho_i\big(\cos\phi_i,\sin\phi_i,0\big)=\rho_i{\vec n}_i.
  \label{eq:coord}
\end{eqnarray}
Defining, for convenience,
{}
\begin{eqnarray}
\eta_i=k\frac{\rho_i}{f},
\label{eq:zerJ2}
\end{eqnarray}
we present the Fresnel--Kirchhoff diffraction formula \cite{Turyshev-Toth:2020-image,Turyshev-Toth:2021-imaging} as
{}
\begin{eqnarray}
{\cal B}({\vec x},{\vec x}_i)&=&
- \frac{e^{ikf(1+{{\vec x}_i^2}/{2f^2})}}{i\lambda f}\iint\displaylimits_{|{\vec x}'|^2\leq (\frac{1}{2}d)^2} d^2{\vec x}'\,
{ B}({\vec x},{\vec x}') e^{-i\eta_i({\vec x}'\cdot{\vec n}_i)}=\nonumber\\
&=&- \frac{e^{ikf(1+{{\vec x}_i^2}/{2f^2})}}{i\lambda f} \frac{1}{{2\pi}}\int_0^{2\pi} d\phi_\xi\int_0^{d/2}\hskip 0pt \rho' d\rho'  \int_0^{2\pi}  \hskip 0pt d\phi '\,\,
 \times \nonumber\\
&&
\hskip 20pt
\times \exp\Big[-i\Big(\alpha\big(\rho\cos(\phi_\xi-\phi)+\rho'\cos(\phi_\xi-\phi')\big)+\beta_2\cos[2(\phi_\xi-\phi_s)]\Big)-i\eta_i\rho'\cos(\phi'-\phi_i)\Big].
  \label{eq:amp-w-f}
\end{eqnarray}

We transform the part of the $\rho$-dependent phase in the expression above as
{}
\begin{eqnarray}
\alpha\rho'\cos(\phi_\xi-\phi')\big)+\eta_i\rho'\cos(\phi_i-\phi')=\rho'\,
u(\phi_\xi,\phi_i) \cos\big(\phi'-\epsilon\big),
  \label{eq:ph4}
\end{eqnarray}
with the quantities $u$ and $\epsilon$ given by the following relationships:
{}
\begin{eqnarray}
u(\phi_\xi,\phi_i)=\sqrt{\alpha^2+2\alpha\eta_i\cos\big(\phi_\xi-\phi_i\big)+\eta_i^2},
\qquad
\cos\epsilon=\frac{\alpha  \cos\phi_\xi+\eta_i\cos\phi_i}{u},
\qquad
\sin\epsilon=\frac{\alpha  \sin\phi_\xi+\eta_i\sin\phi_i}{u}.
  \label{eq:eps}
\end{eqnarray}

Thus, we transform the expression given by (\ref{eq:amp-w-f}) as
{}
\begin{eqnarray}
{\cal B}({\vec x},{\vec x}_i)&=&ie^{ikf(1+{{\vec x}_i^2}/{2f^2})}
\Big(\frac{kd^2}{8f}\Big) \times\nonumber\\
&&\times\frac{1}{{2\pi}}\int_0^{2\pi} d\phi_\xi \Big(\frac{2J_1\big(u(\phi_\xi,\phi_i){\textstyle\frac{1}{2}}d\big)}{u(\phi_\xi,\phi_i){\textstyle\frac{1}{2}}d}\Big)
\exp\Big[-i\Big(\alpha\rho\cos(\phi_\xi-\phi)+\beta_2\cos[2(\phi_\xi-\phi_s)]\Big)\Big],
  \label{eq:amp-dsf}
\end{eqnarray}
which may be evaluated with the method of stationary phase using the quartic solution.

Next, with the amplitude ${\cal B}({\vec x},{\vec x}_i)$ given by (\ref{eq:amp-dsf}), the EM field in the focal plane of the telescope (indicated by subscript ${\vec x}_i$) produced by a point source using (\ref{eq:DB-sol-rho}) is given as
{}
\begin{eqnarray}
    \left( \begin{aligned}
{E}_\rho& \\
{H}_\rho& \\
  \end{aligned} \right)_{\hskip -3pt {\vec x}_i} =    \left( \begin{aligned}
{H}_\phi& \\
-{E}_\phi& \\
  \end{aligned} \right)_{\hskip -3pt \vec x_i} &=&E_0
  \sqrt{2\pi kr_g} e^{i\sigma_0}  {\cal B}({\vec x},{\vec x}_i)\, e^{i(kz-\omega t)}
 \left( \begin{aligned}
 \cos\phi& \\
 \sin\phi& \\
  \end{aligned} \right).
  \label{eq:DB-sol-rho2}
\end{eqnarray}

With overline and brackets denoting time-averaging and ensemble averaging
(over the source's surface), correspondingly, and defining $\Omega(t)=kz-\omega t$, we compute $S_z$ as
 {}
\begin{eqnarray}
S_z({\vec x},{\vec x}_i)=\frac{c}{4\pi}\big<\overline{[{\rm Re}{\vec E}\times{\rm Re}{\vec H}]}_z\big>=\frac{c}{4\pi} E_0^2\,{2\pi kr_g}\,
\big<\overline{\big({\rm Re}\big[{\cal B}({\vec x},{\vec x}_i)e^{i\Omega(t)}\big]\big)^2}\big>.
  \label{eq:Pv}
\end{eqnarray}

Using the expression for the Poynting vector carried by an EM wave in a vacuum in flat spacetime and seen on the focal plane of the imaging telescope, $\vec S_0({\vec x},{\vec x}_i)$, we can now obtain the light amplification factor, $\mu({\vec x},{\vec x}_i)=S_z({\vec x},{\vec x}_i)/|{\vec S}_0|$  of the optical system consisting of the SGL and an imaging telescope, i.e., the convolution of the PSF of the SGL with that of an optical telescope. From (\ref{eq:Pv}), this expression takes the form, $\mu({\vec x},{\vec x}_i)=2 \big<\overline{\big({\rm Re}\big[{\cal B}({\vec x},{\vec x}_i)e^{i\Omega(t)}\big]\big)^2}\big>.$ As a result, we have the light amplification factor of the lens that, for short wavelengths (i.e., $kr_g\gg1$) is given as
 {}
\begin{eqnarray}
\mu_z({\vec x},{\vec x}_i)={2\pi kr_g}  \, I({\vec x},{\vec x}_i),
\qquad {\rm where}\qquad I({\vec x},{\vec x}_i)=|{\cal B}({\vec x},{\vec x}_i)|^2,
  \label{eq:S_mu}
\end{eqnarray}
with ${\cal B}^*({\vec x},{\vec x}_i)$ is the complex conjugate of ${\cal B}({\vec x},{\vec x}_i)$, and $I({\vec x},{\vec x}_i)$ is the intensity of EM field deposited on the focal plane of the imaging telescope when imaging with the SGL (see details in \cite{Turyshev-Toth:2021-imaging}).

\subsection{Using the quartic solution to evaluate the intensity}

\begin{figure}
\begin{tabular}{>{\centering}p{0.4\textwidth} >{\centering}p{0.2\textwidth} >{\centering}p{0.2\textwidth} >{\centering}p{0.2\textwidth}}
Quartic roots&Telescope position&Quartic lens view&Numerical integration \cr
\vspace{-0.15in} \includegraphics{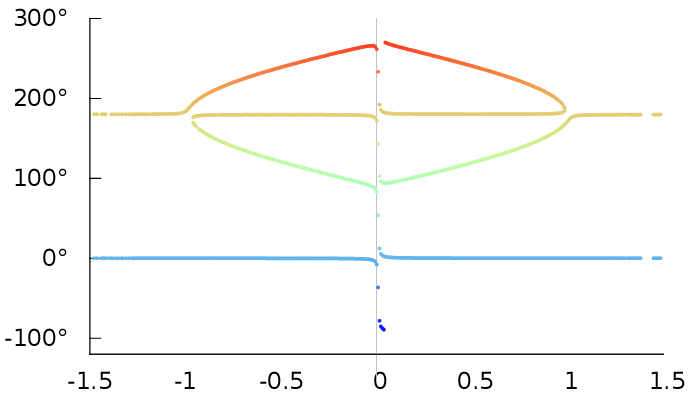} &
\vspace{0.1in} \includegraphics[scale=0.5]{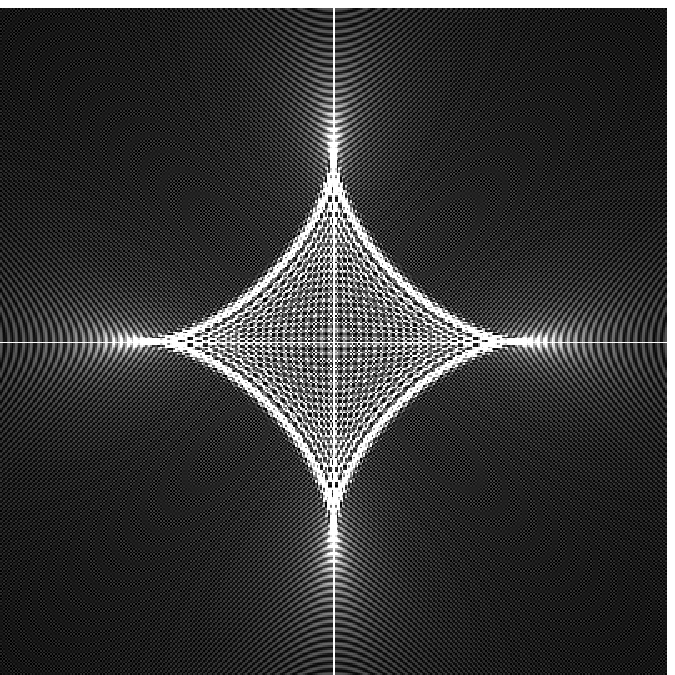} &
\vspace{0.1in} \includegraphics[scale=0.5]{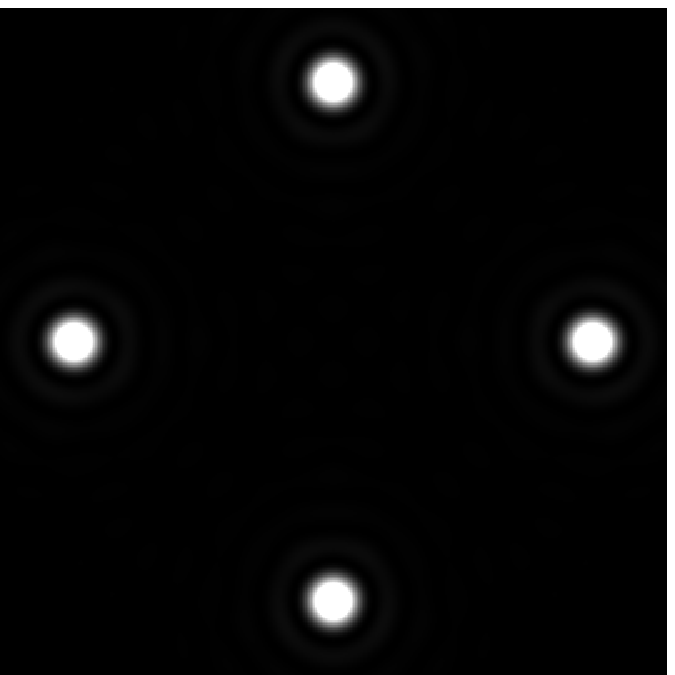} &
\vspace{0.1in} \includegraphics[scale=0.5]{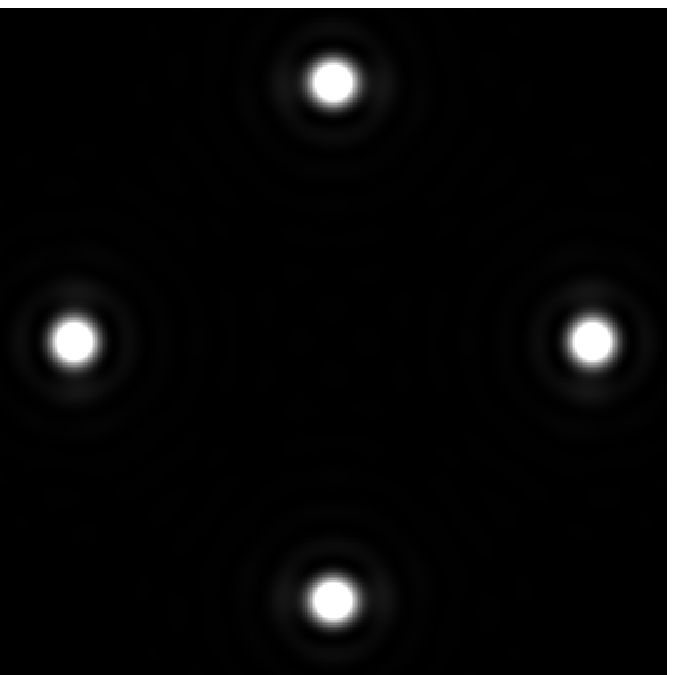} \cr
\vspace{-0.15in} \includegraphics{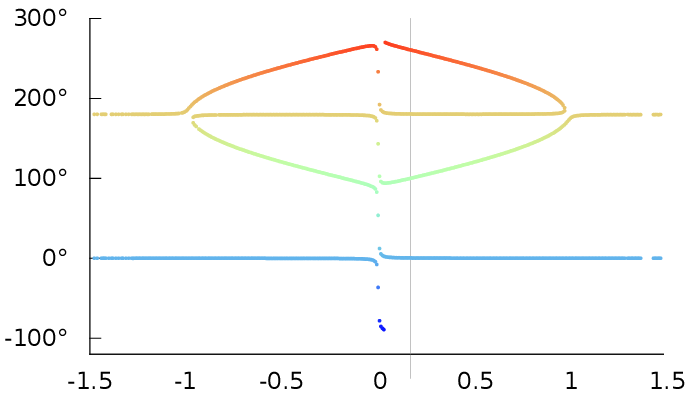} &
\vspace{0.1in} \includegraphics[scale=0.5]{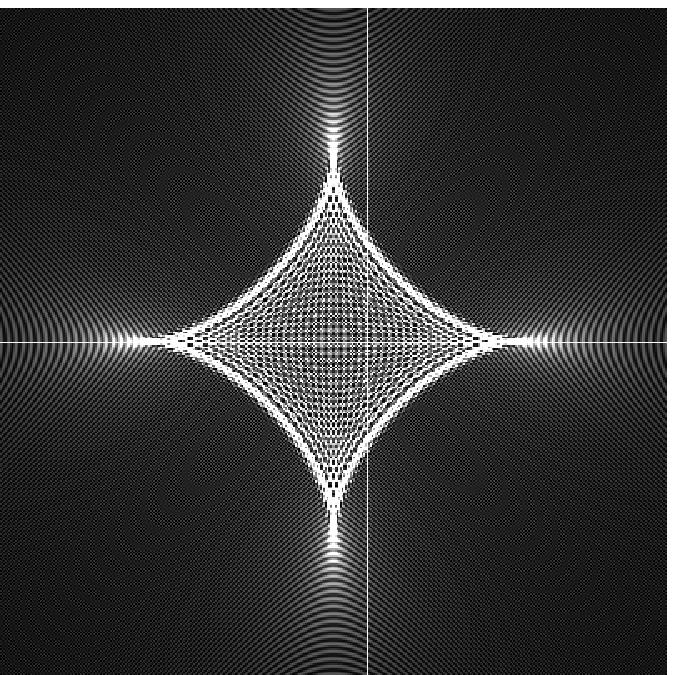} &
\vspace{0.1in} \includegraphics[scale=0.5]{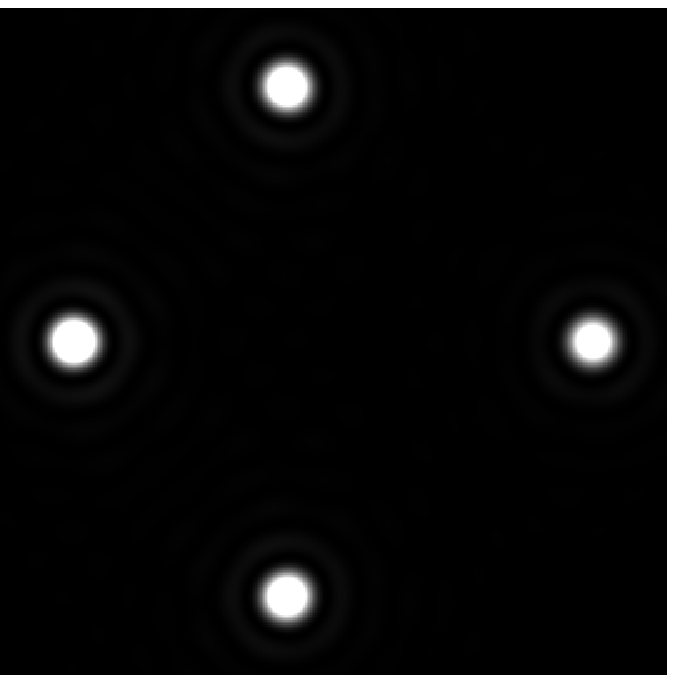} &
\vspace{0.1in} \includegraphics[scale=0.5]{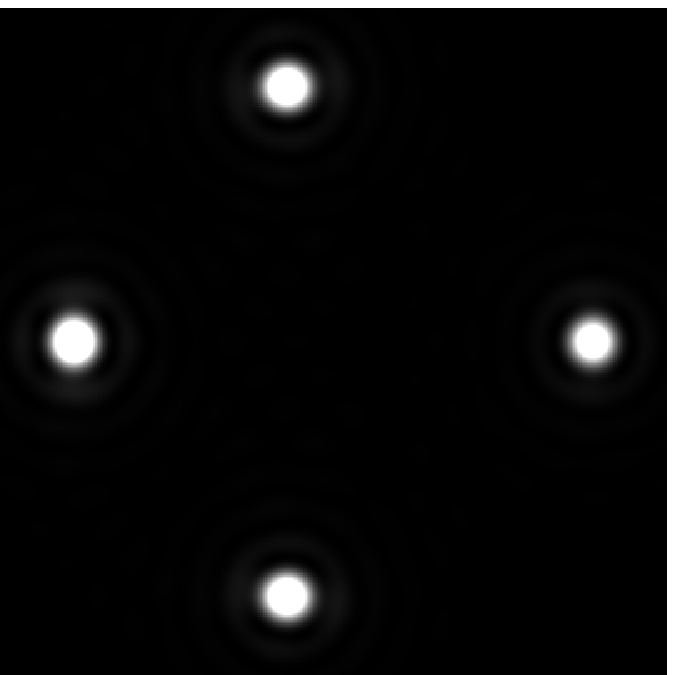} \cr
\vspace{-0.15in} \includegraphics{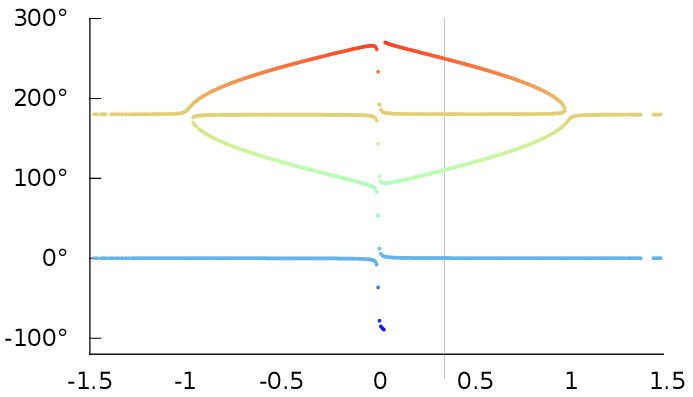} &
\vspace{0.1in} \includegraphics[scale=0.5]{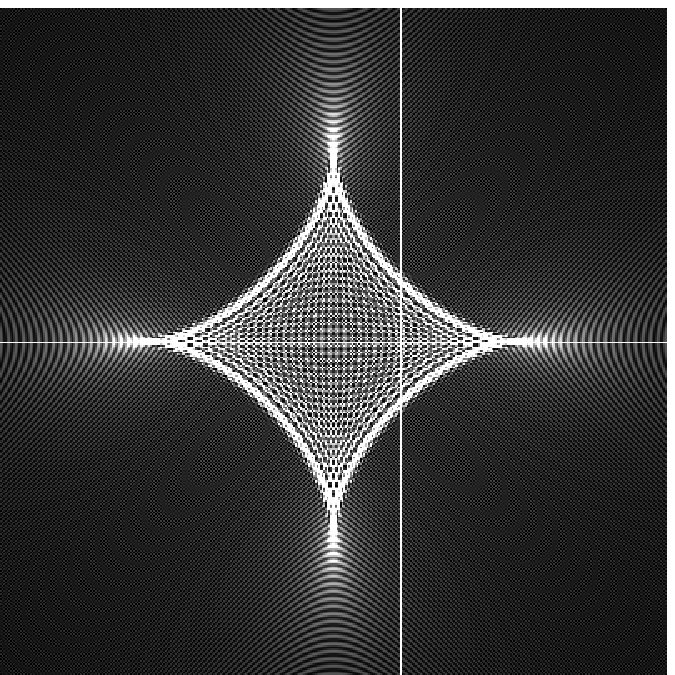} &
\vspace{0.1in} \includegraphics[scale=0.5]{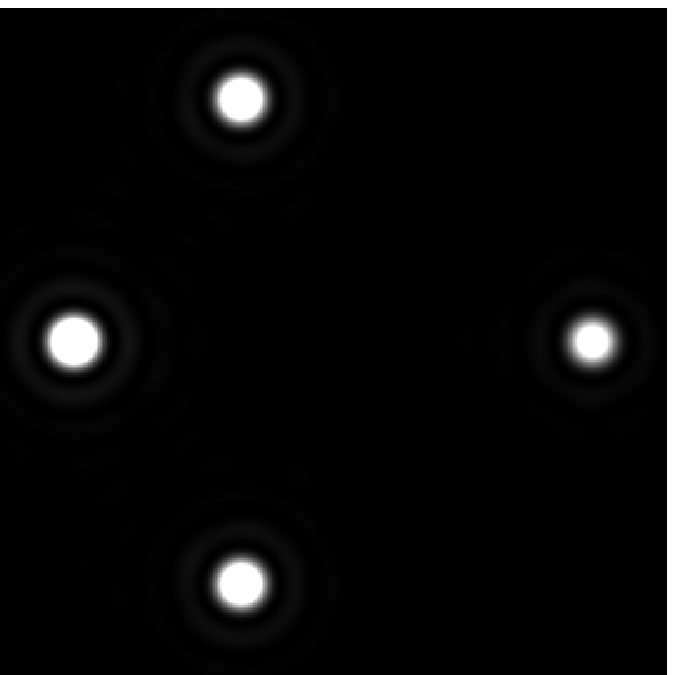} &
\vspace{0.1in} \includegraphics[scale=0.5]{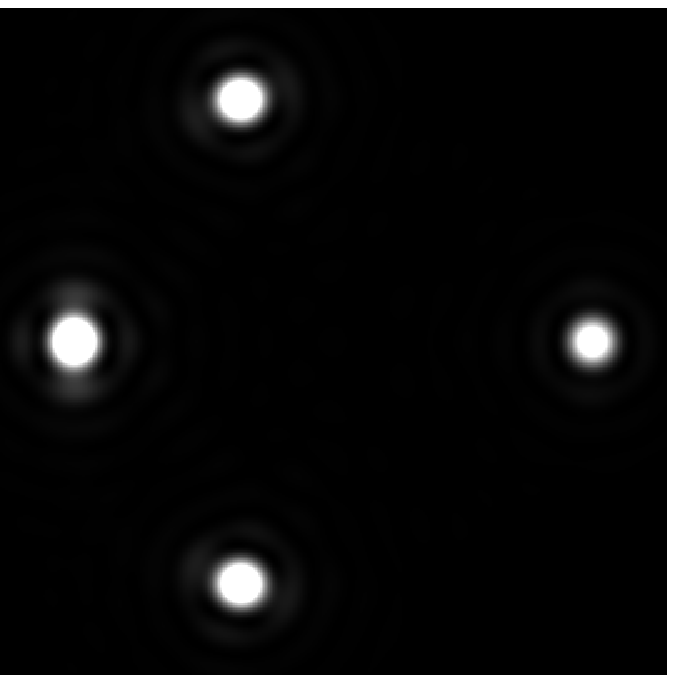} \cr
\vspace{-0.15in} \includegraphics{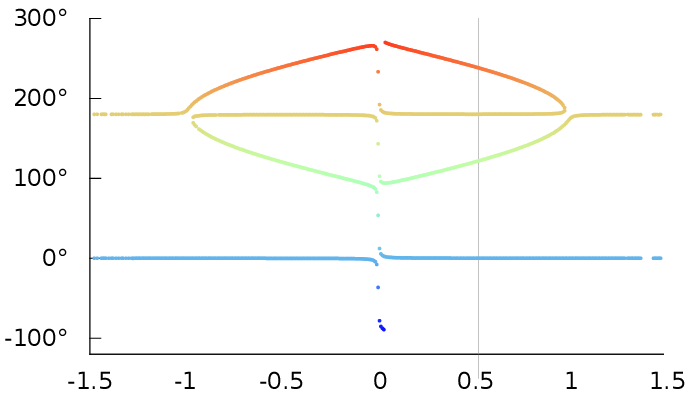} &
\vspace{0.1in} \includegraphics[scale=0.5]{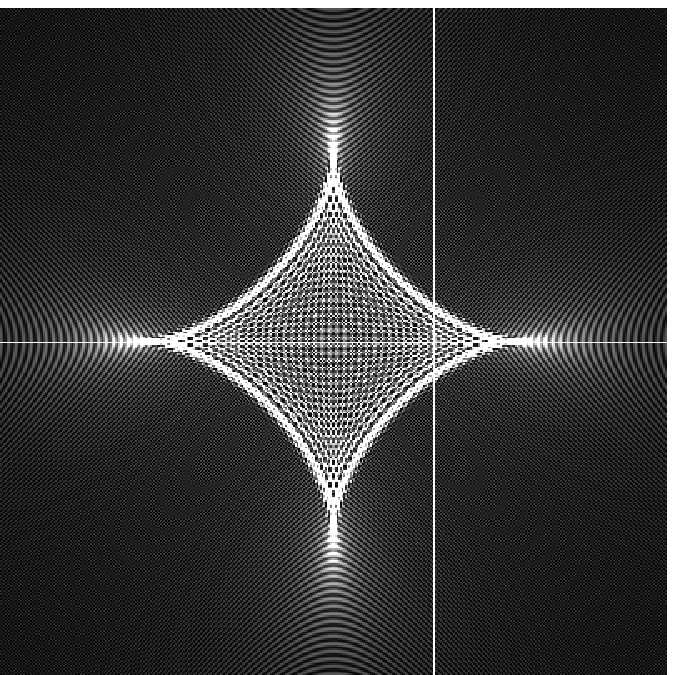} &
\vspace{0.1in} \includegraphics[scale=0.5]{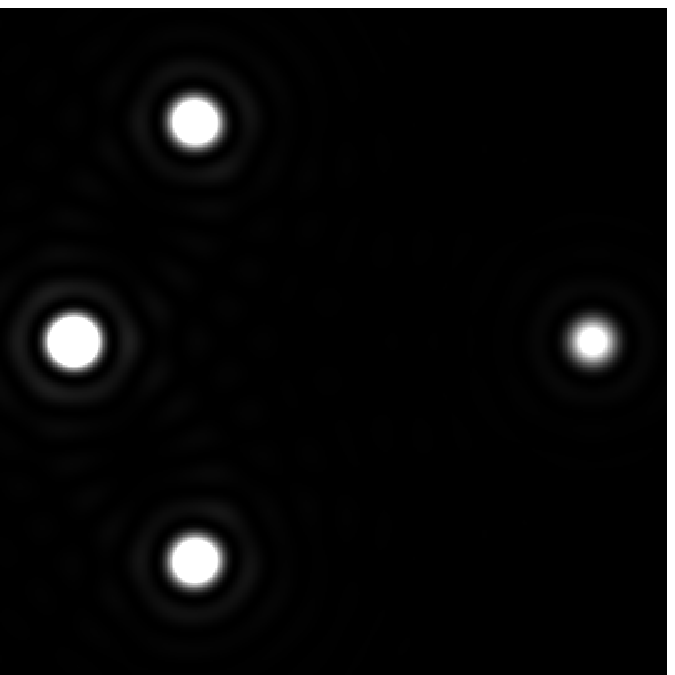} &
\vspace{0.1in} \includegraphics[scale=0.5]{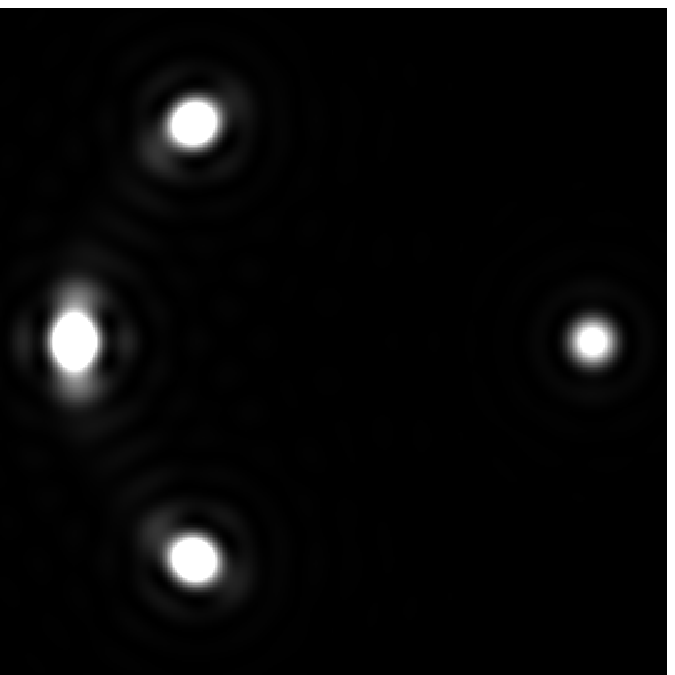}
\end{tabular}
\caption{\label{fig:cross-0-3}Studying the quartic solution in the context of views seen by the imaging telescope. The first column shows the telescope position (thin vertical gray line) overlayed the cross-sectional view of the quartic solution from Fig.~\ref{fig:crossdir}, with $\eta$ on the horizontal axis and $(\phi_\xi-\phi)$, given in degrees, on the vertical. Second column: Crosshairs indicating the same telescope location with respect to the PSF. Third column: the view through the telescope, modeled using the quartic solution. Fourth column: Telescopic view via numerical integration. The parametrization is that of the Sun at 650~AU, with $\sin\beta_s=0.02$, such that $\eta=1$ corresponds to approximately 5.6 meters. The telescope is positioned in 1~m increments from 0 to 3~m. We see that the azimuthal positions of the four spots of light in the Einstein cross seen by the telescope correspond exactly to the quartic roots; these roots, however, do not model the widening of these light spots into partial arcs, seen in the full numerical evaluation.}
\end{figure}

\begin{figure}
\begin{tabular}{>{\centering}p{0.4\textwidth} >{\centering}p{0.2\textwidth} >{\centering}p{0.2\textwidth} >{\centering}p{0.2\textwidth}}
Quartic roots&Telescope position&Quartic lens view&Numerical integration \cr
\vspace{-0.15in} \includegraphics{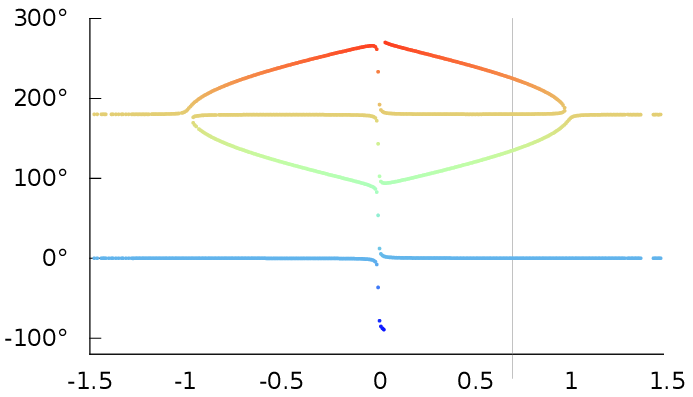} &
\vspace{0.1in} \includegraphics[scale=0.5]{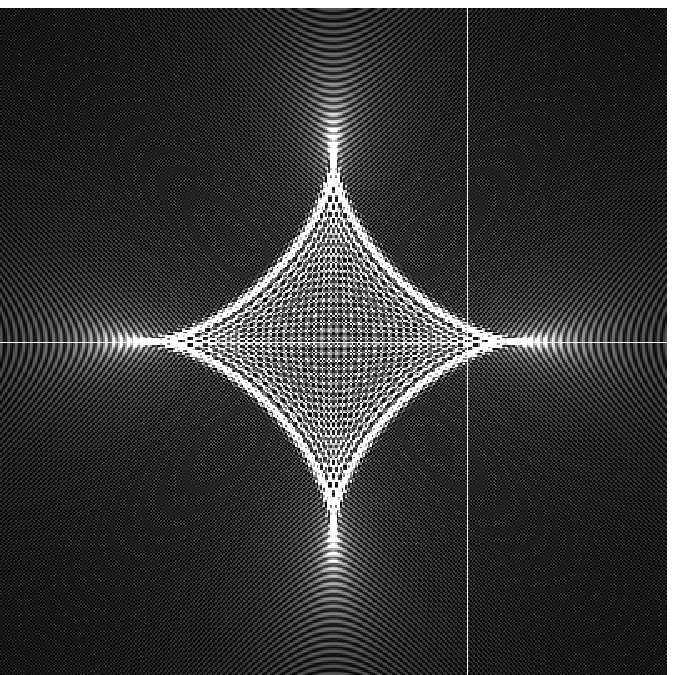} &
\vspace{0.1in} \includegraphics[scale=0.5]{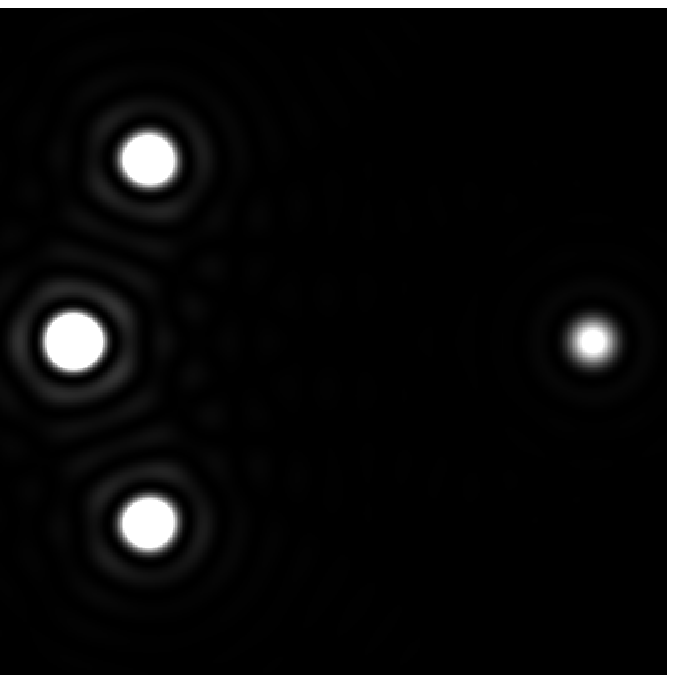} &
\vspace{0.1in} \includegraphics[scale=0.5]{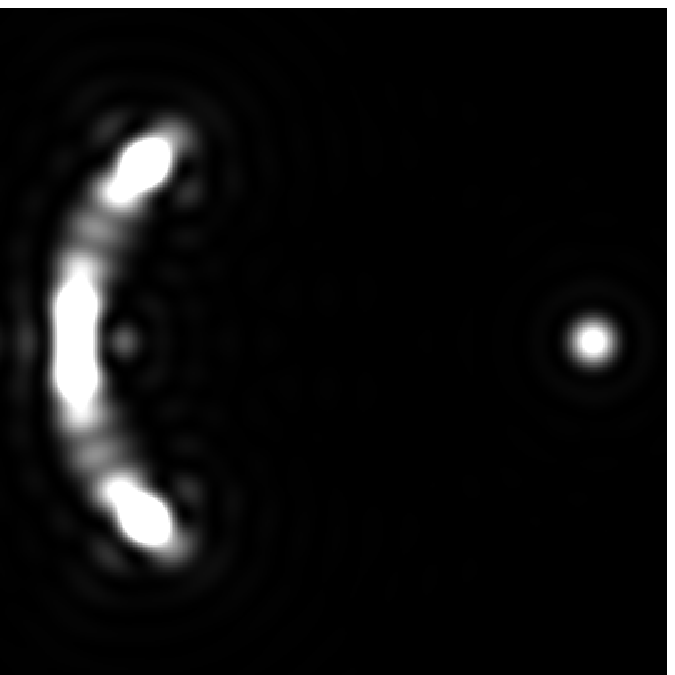} \cr
\vspace{-0.15in} \includegraphics{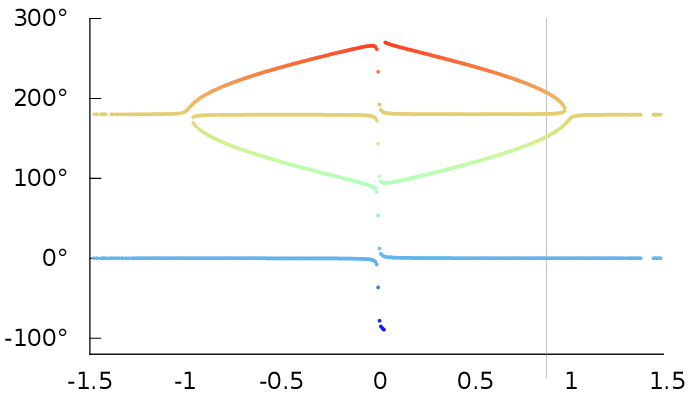} &
\vspace{0.1in} \includegraphics[scale=0.5]{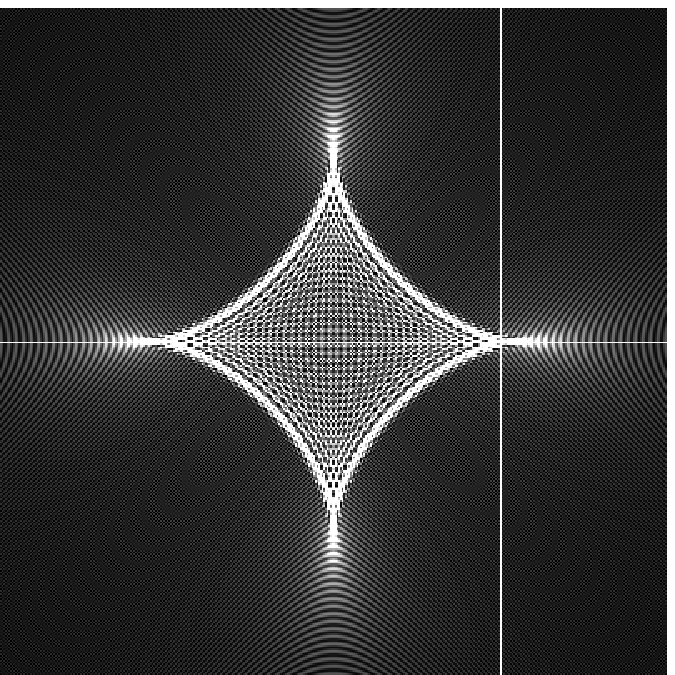} &
\vspace{0.1in} \includegraphics[scale=0.5]{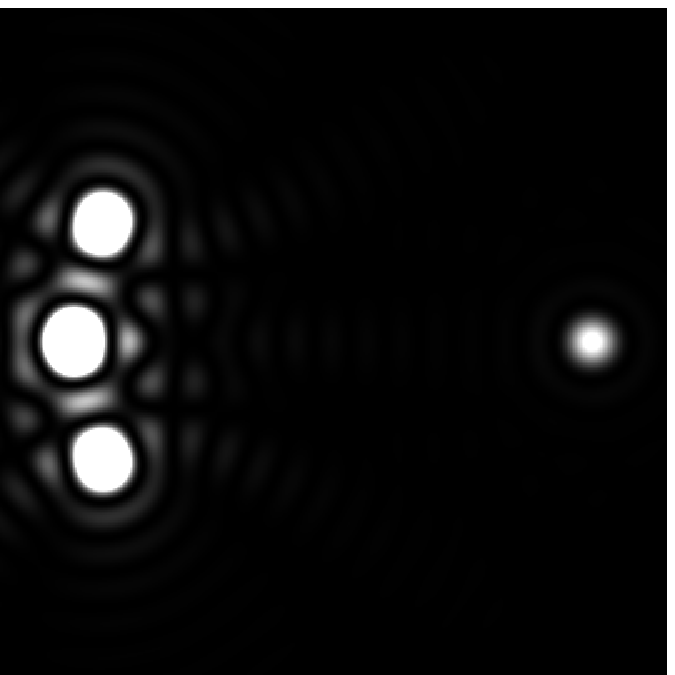} &
\vspace{0.1in} \includegraphics[scale=0.5]{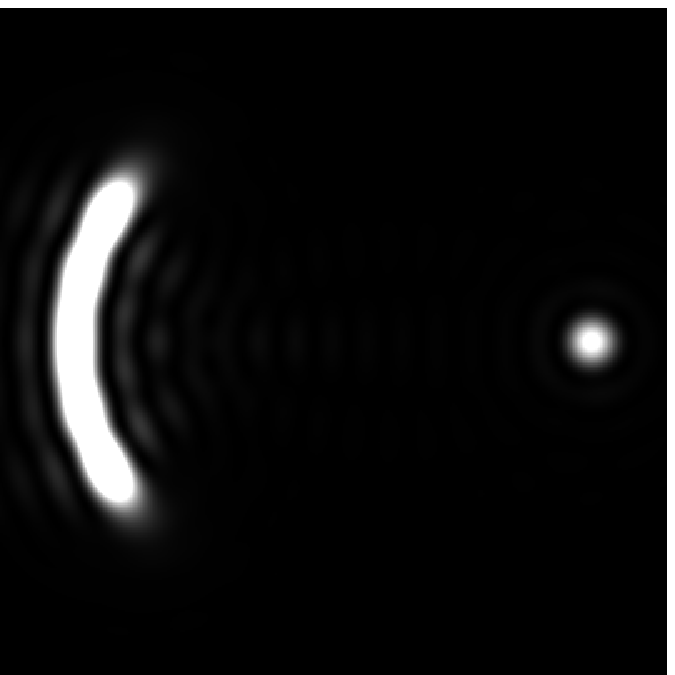} \cr
\vspace{-0.15in} \includegraphics{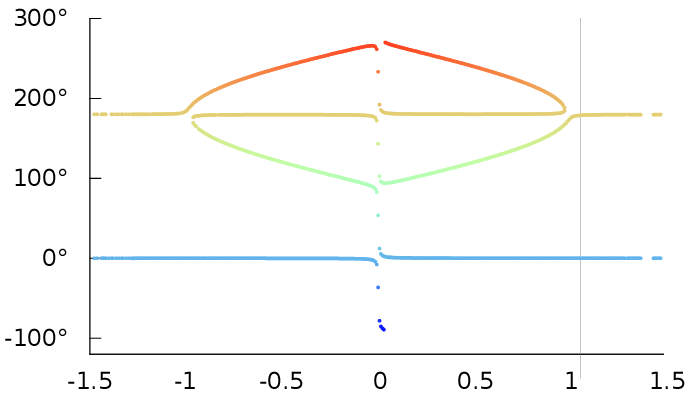} &
\vspace{0.1in} \includegraphics[scale=0.5]{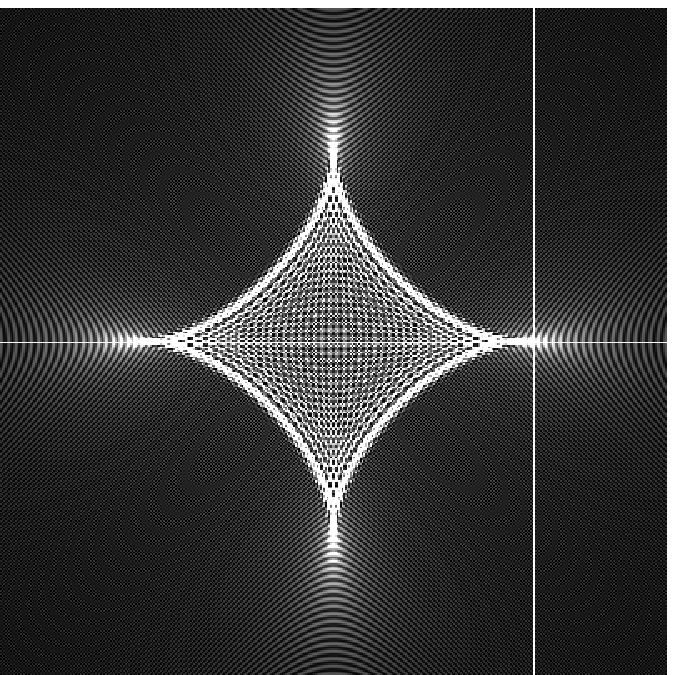} &
\vspace{0.1in} \includegraphics[scale=0.5]{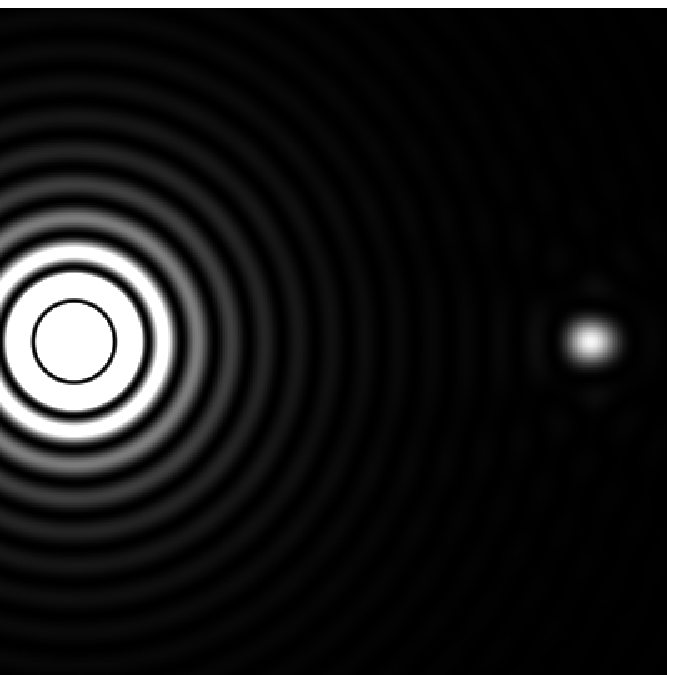} &
\vspace{0.1in} \includegraphics[scale=0.5]{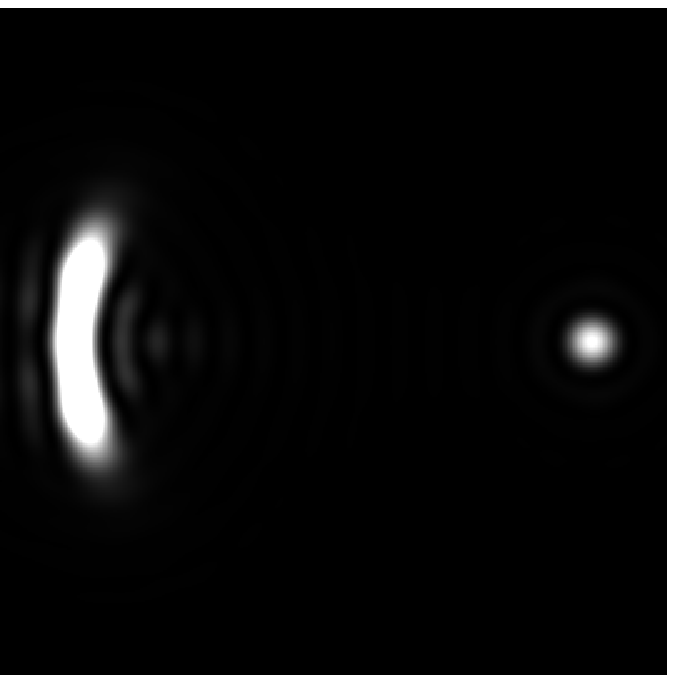} \cr
\vspace{-0.15in} \includegraphics{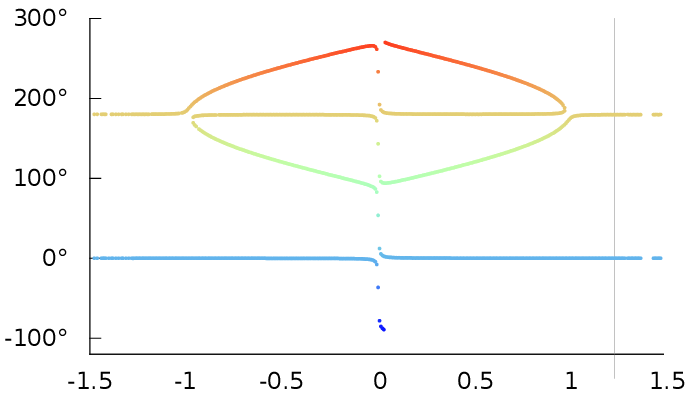} &
\vspace{0.1in} \includegraphics[scale=0.5]{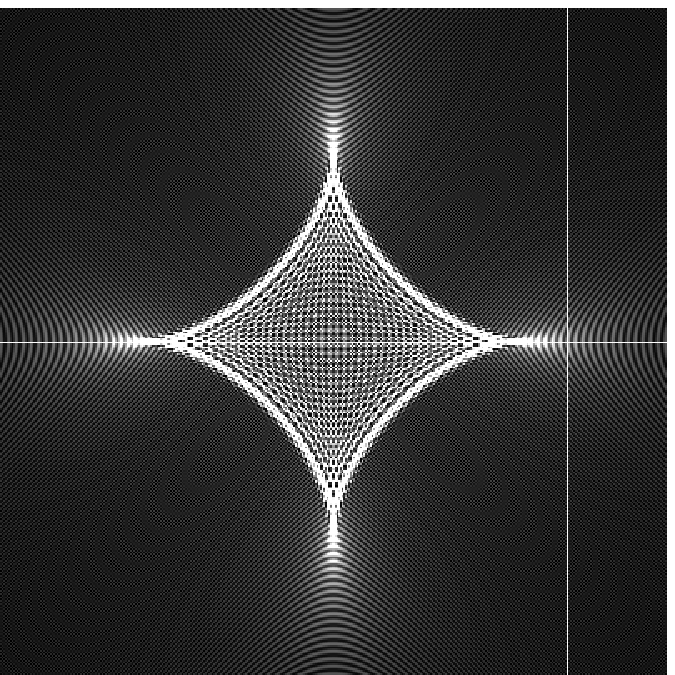} &
\vspace{0.1in} \includegraphics[scale=0.5]{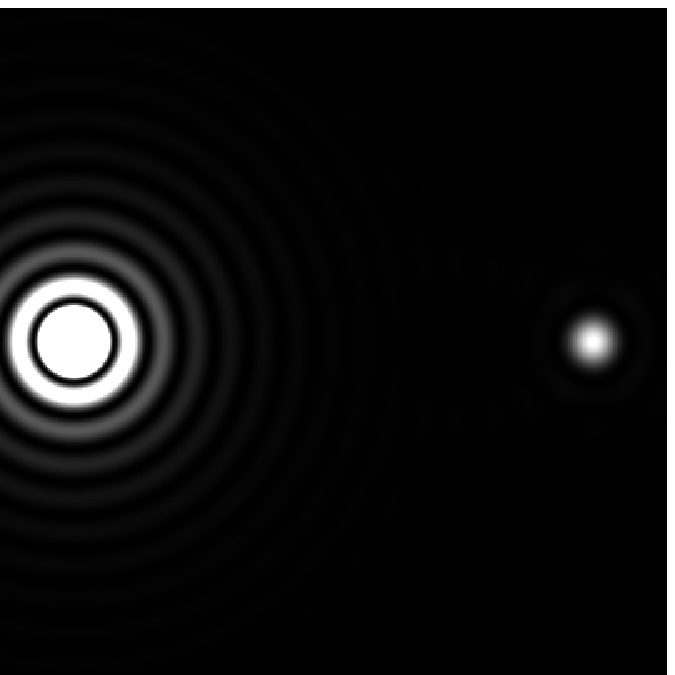} &
\vspace{0.1in} \includegraphics[scale=0.5]{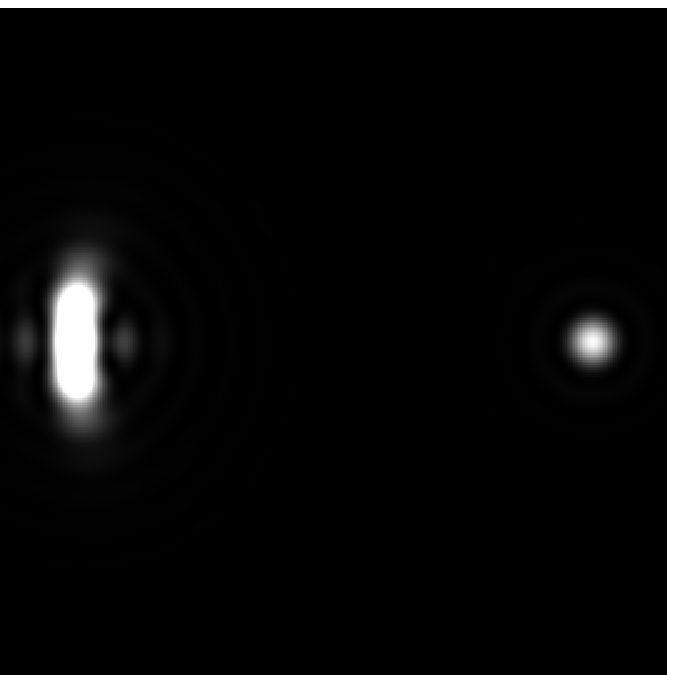}
\end{tabular}
\caption{\label{fig:cross-4-7}As in Fig.~\ref{fig:cross-0-3}, with the telescope now positioned between 4 and 7 meters, in 1~m increments. As the telescope traverses the cusp of the PSF, numerical integration shows how three of the four spots of the Einstein cross merge into an arc that eventually shortens and transitions into a single spot of light. This transition is not captured well by the quartic solution; the three spots of light remain discrete spots until they merge into a single, bright spot which then fades.}
\end{figure}

\begin{figure}
\begin{tabular}{>{\centering}p{0.4\textwidth} >{\centering}p{0.2\textwidth} >{\centering}p{0.2\textwidth} >{\centering}p{0.2\textwidth}}
Quartic roots&Telescope position&Quartic lens view&Numerical integration \cr
\vspace{-0.15in} \includegraphics{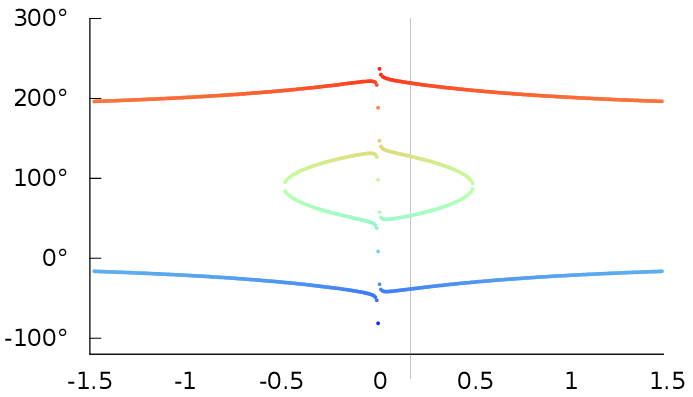} &
\vspace{0.1in} \includegraphics[scale=0.5]{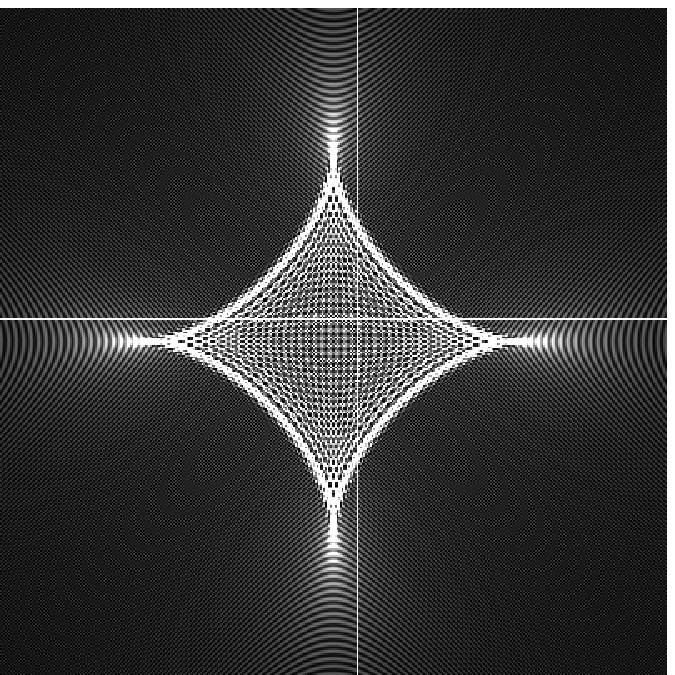} &
\vspace{0.1in} \includegraphics[scale=0.5]{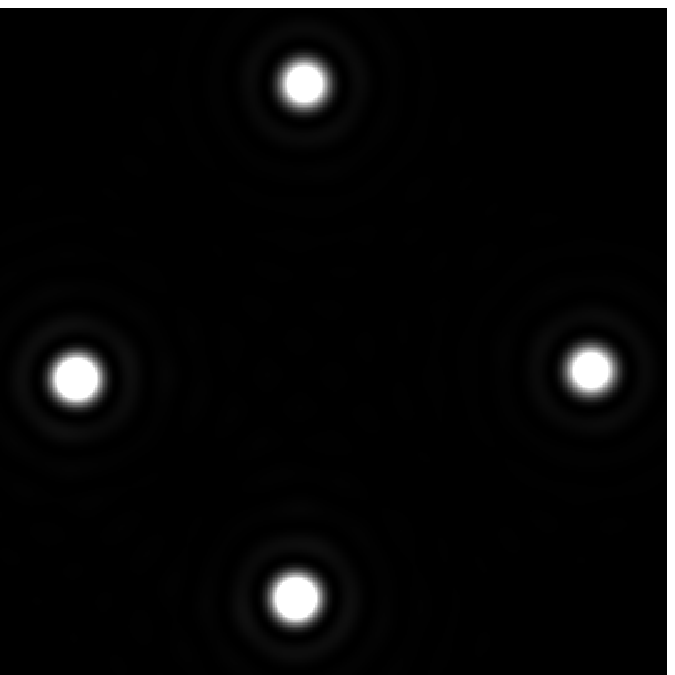} &
\vspace{0.1in} \includegraphics[scale=0.5]{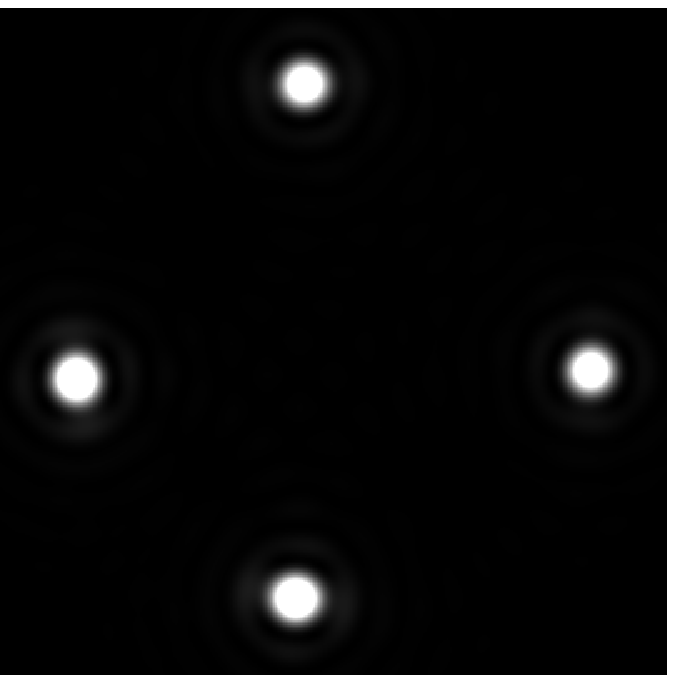} \cr
\vspace{-0.15in} \includegraphics{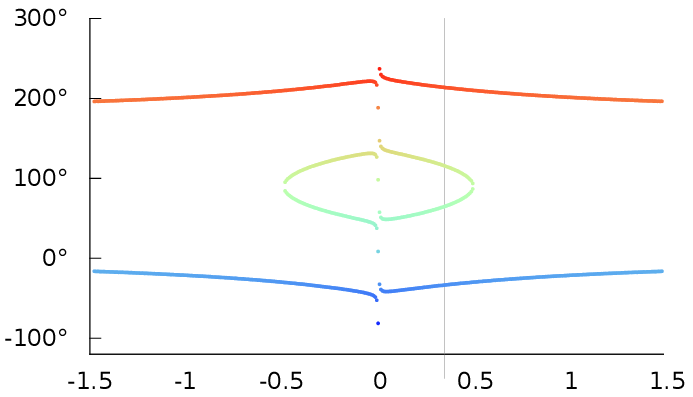} &
\vspace{0.1in} \includegraphics[scale=0.5]{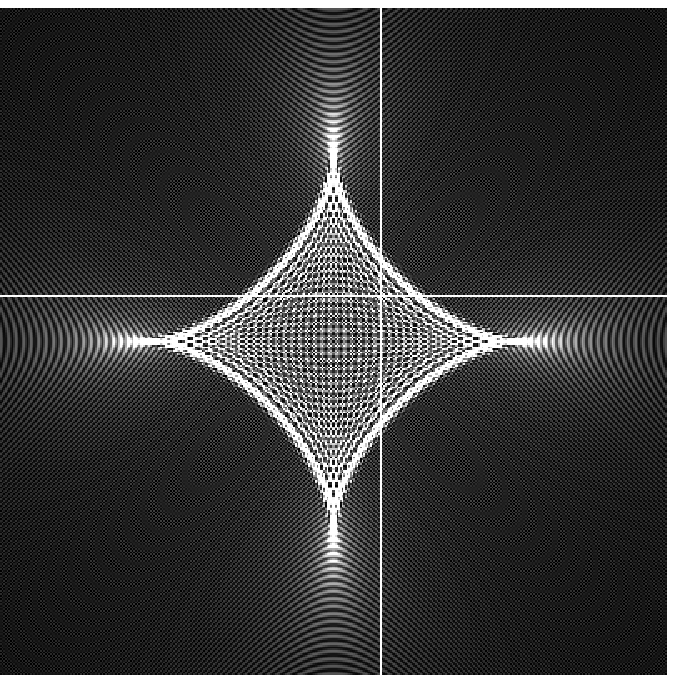} &
\vspace{0.1in} \includegraphics[scale=0.5]{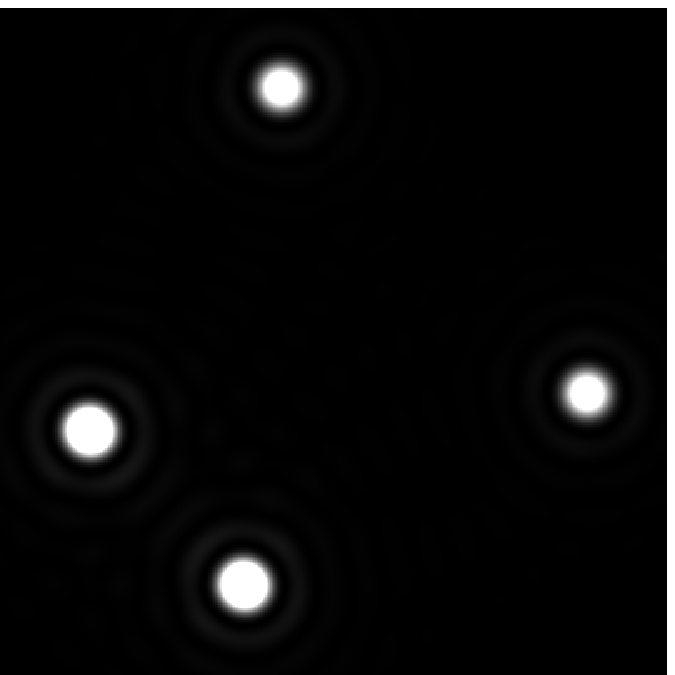} &
\vspace{0.1in} \includegraphics[scale=0.5]{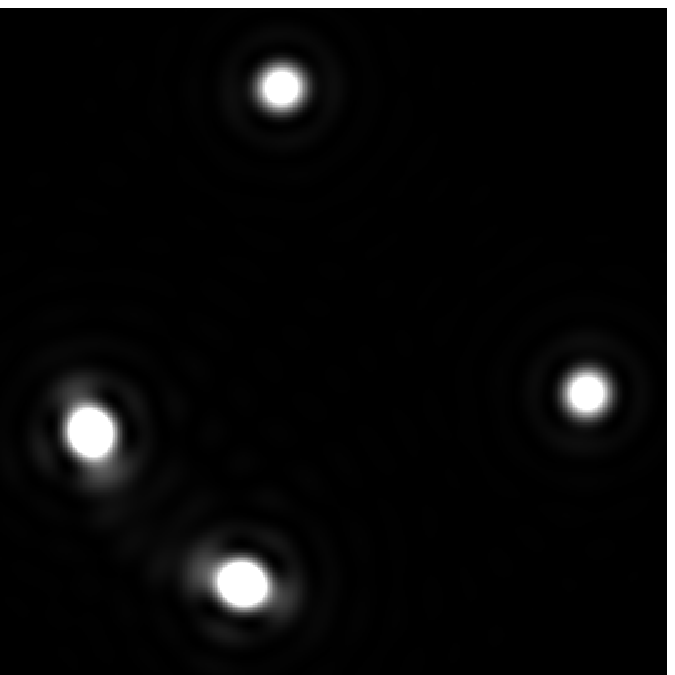} \cr
\vspace{-0.15in} \includegraphics{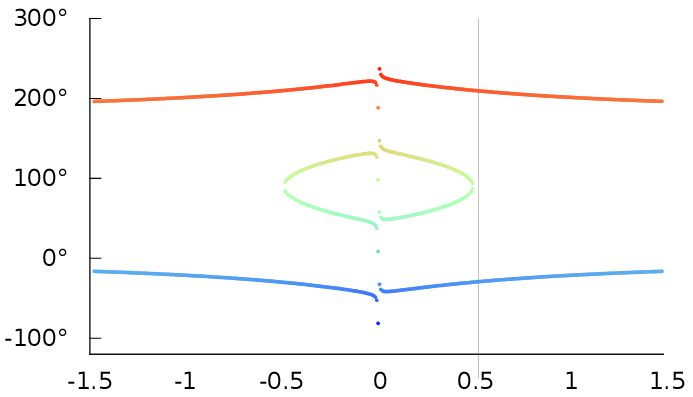} &
\vspace{0.1in} \includegraphics[scale=0.5]{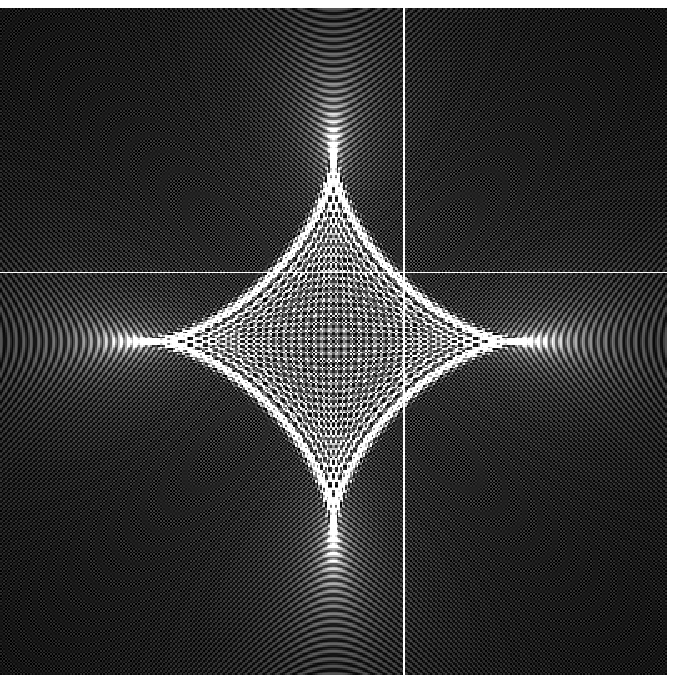} &
\vspace{0.1in} \includegraphics[scale=0.5]{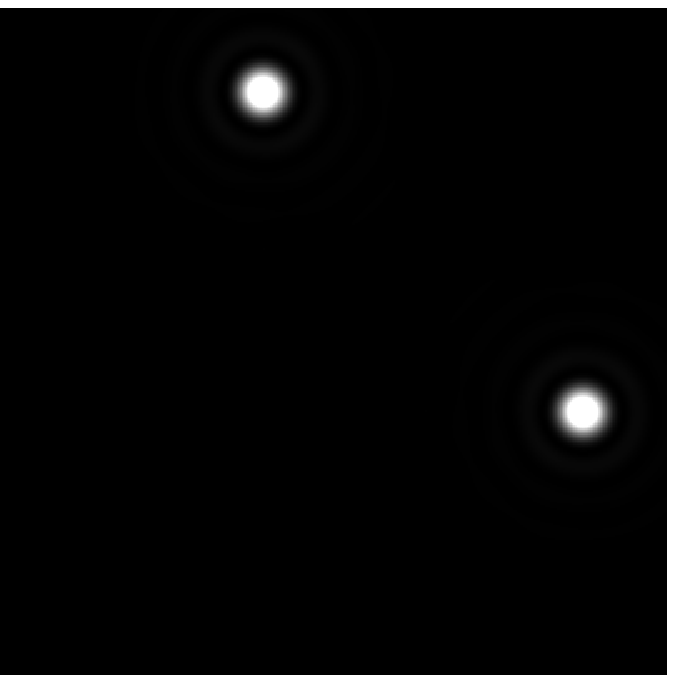} &
\vspace{0.1in} \includegraphics[scale=0.5]{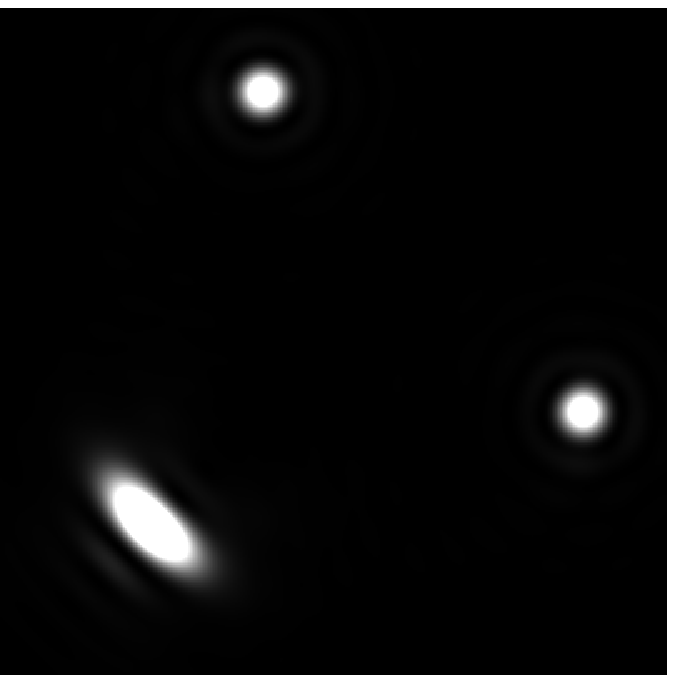} \cr
\vspace{-0.15in} \includegraphics{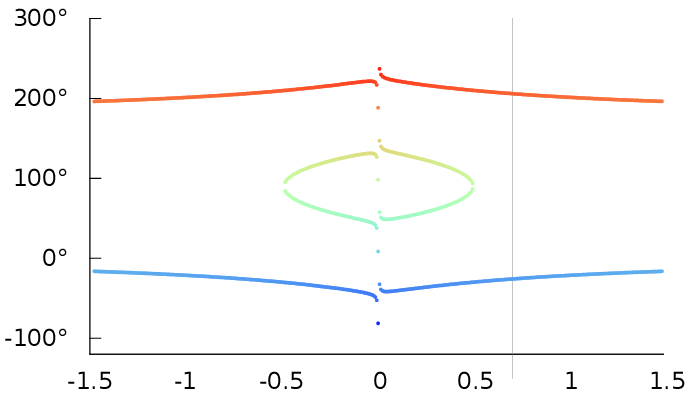} &
\vspace{0.1in} \includegraphics[scale=0.5]{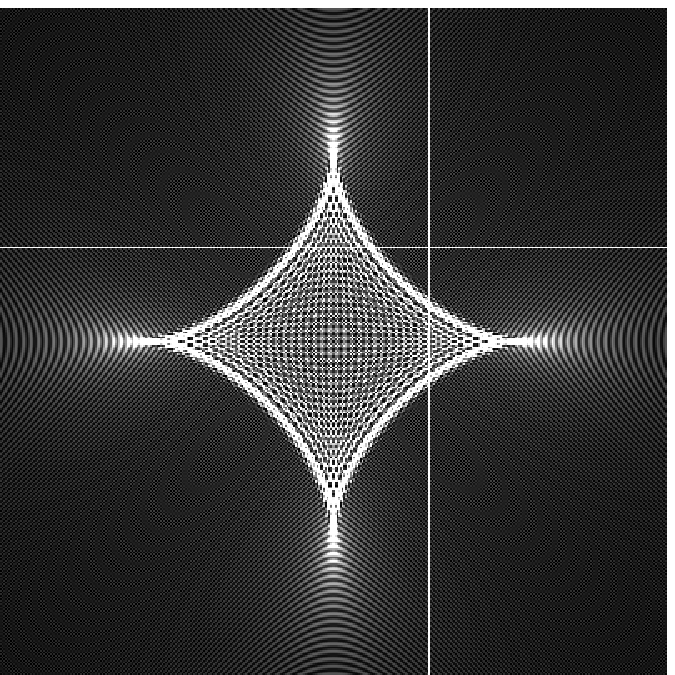} &
\vspace{0.1in} \includegraphics[scale=0.5]{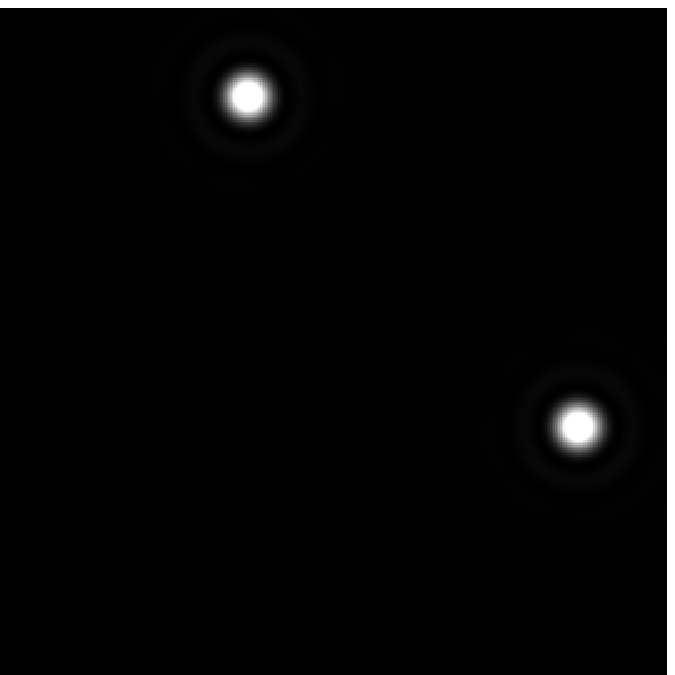} &
\vspace{0.1in} \includegraphics[scale=0.5]{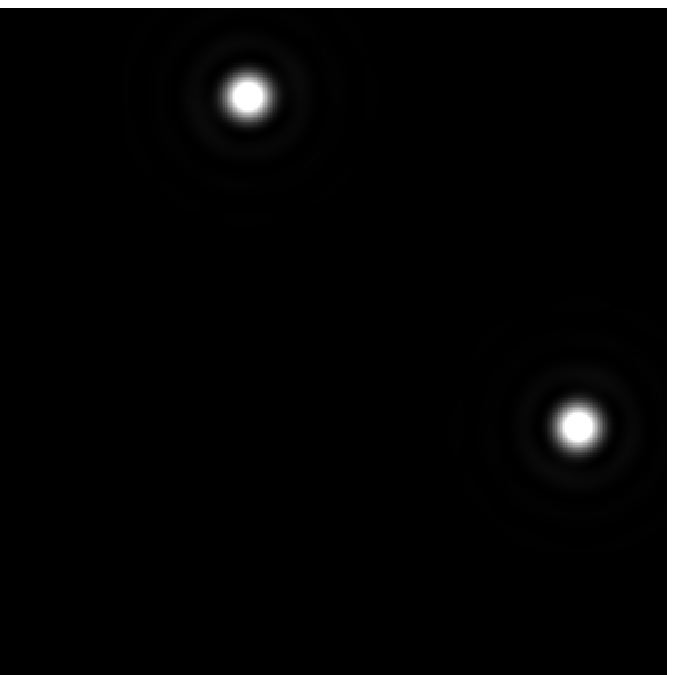}
\end{tabular}
\caption{\label{fig:fold-1-4}As in Fig.~\ref{fig:cross-0-3}, but the telescope is now moving in the fold direction, $\tfrac{1}{2}\mu=\pi/4$, positioned between 1 and 4 meters, in 1~m increments. Again we see how the transition from the interior region of the caustic to the exterior is represented by a change from four to two roots in the quartic equation. The transition in the quartic model is sudden, corresponding to the sharp caustic boundary we saw earlier in Fig.~\ref{fig:astrdet} (left), as two of the four spots of light merge and vanish. Numerical integration reveals a more gradual transition as the merging spots form a short arc that shortens and fades.}
\end{figure}

\begin{figure}
\begin{tabular}{>{\centering}p{0.4\textwidth} >{\centering}p{0.4\textwidth} >{\centering}p{0.2\textwidth}}
Polynomial&Quartic roots&Quartic lens view \cr
\vspace{-0.115in} \includegraphics{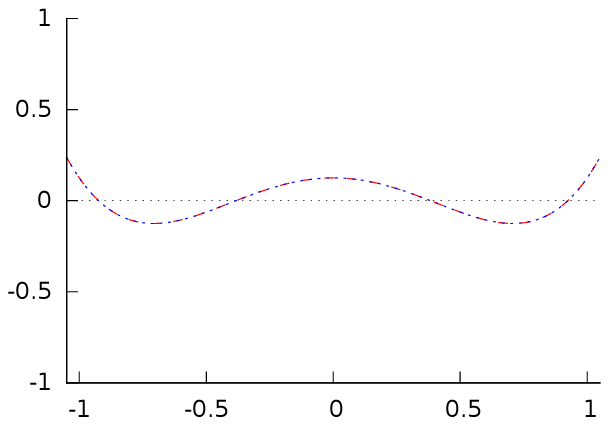} &
\vspace{-0.15in} \includegraphics{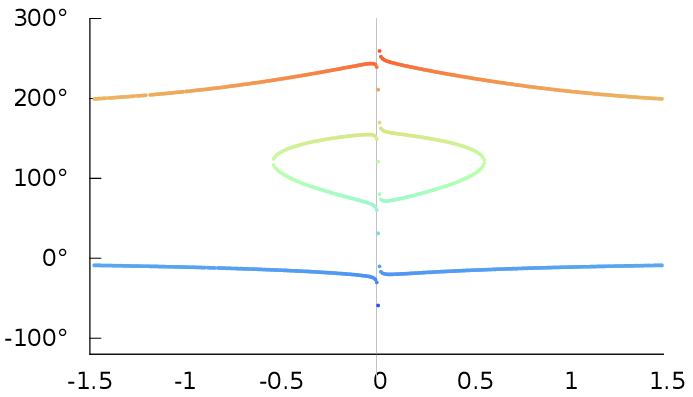} &
\vspace{0.1in} \includegraphics[scale=0.5]{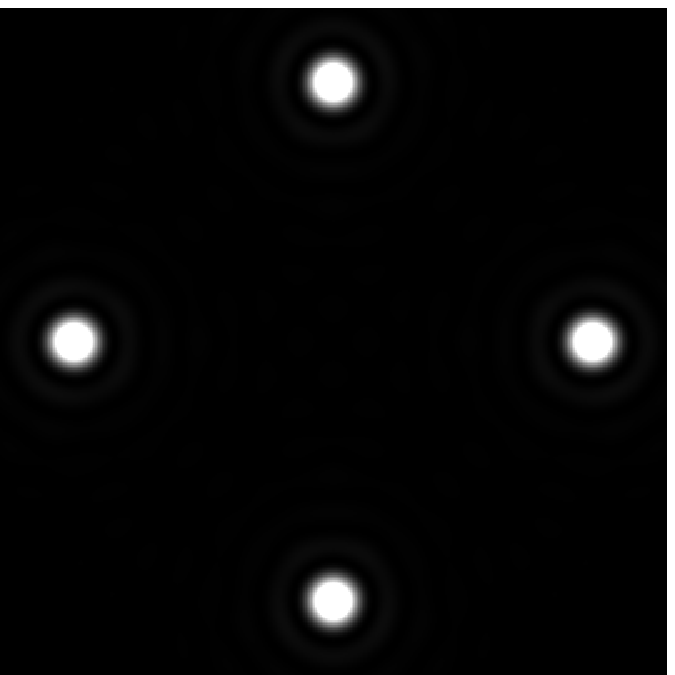} \cr
\vspace{-0.115in} \includegraphics{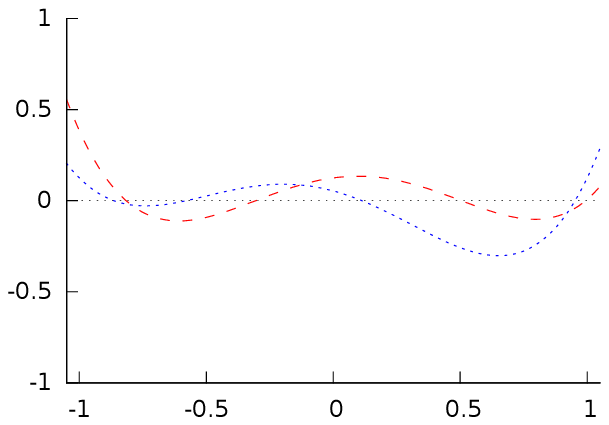} &
\vspace{-0.15in} \includegraphics{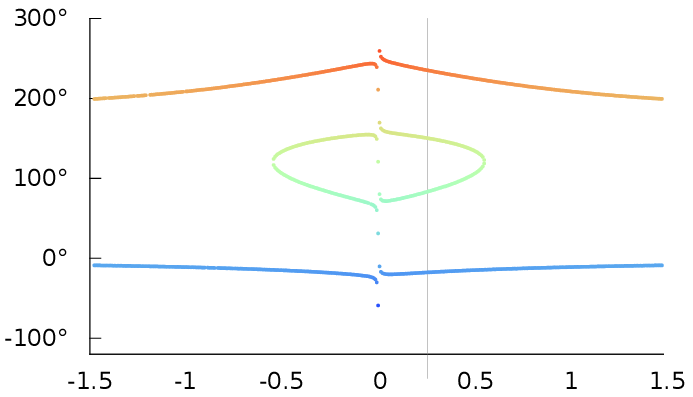} &
\vspace{0.1in} \includegraphics[scale=0.5]{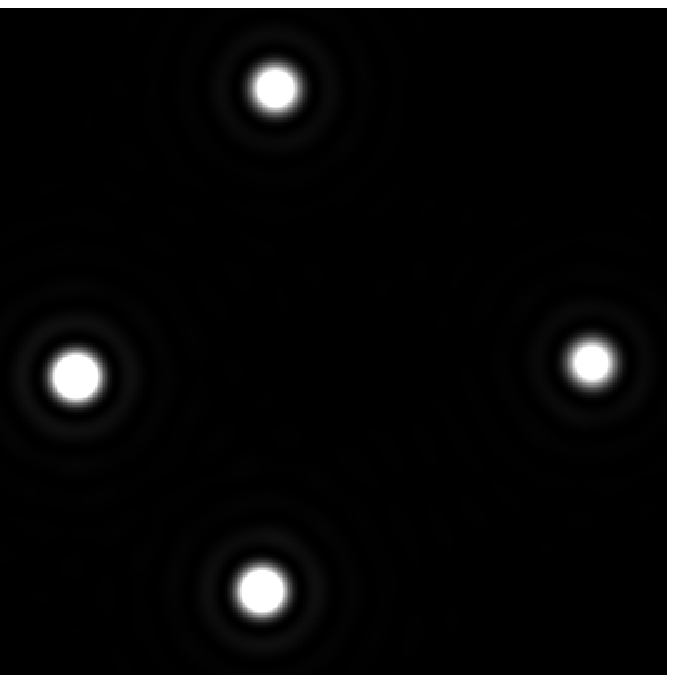} \cr
\vspace{-0.115in} \includegraphics{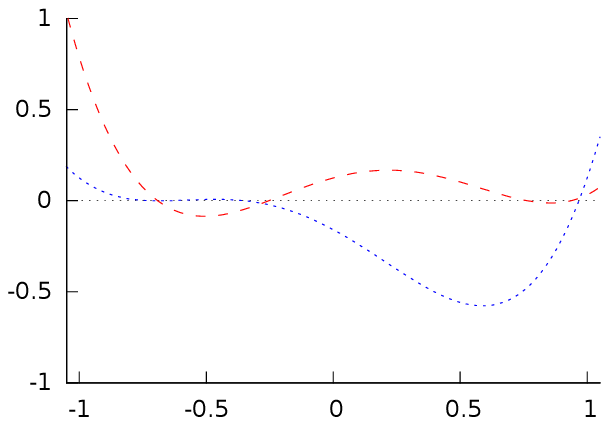} &
\vspace{-0.15in} \includegraphics{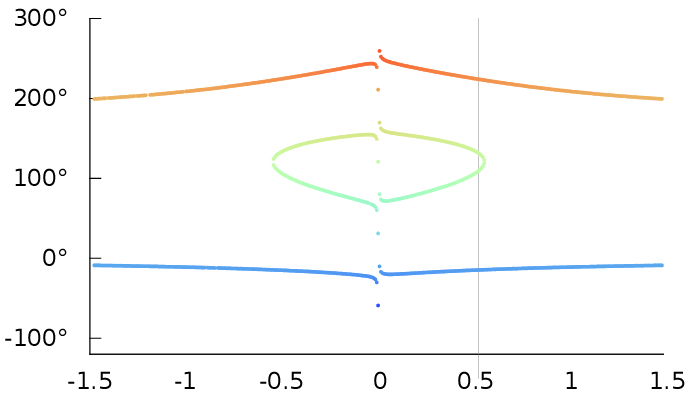} &
\vspace{0.1in} \includegraphics[scale=0.5]{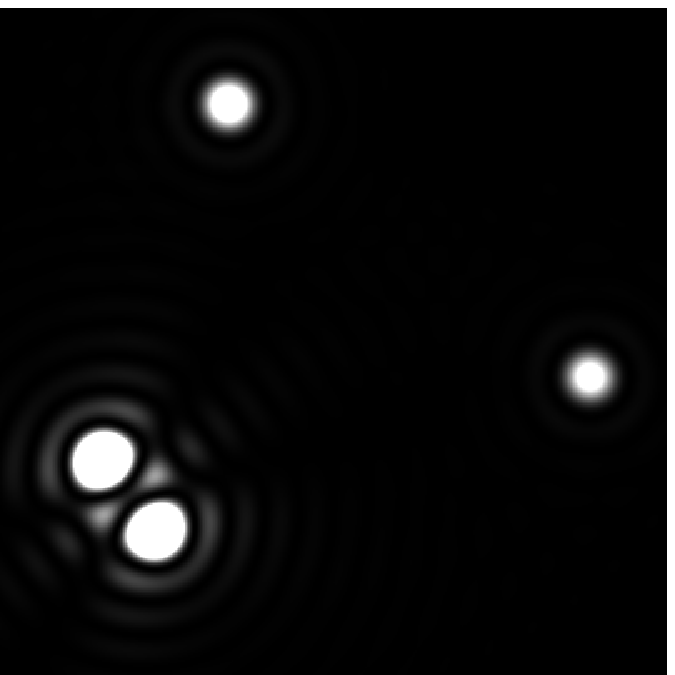} \cr
\vspace{-0.115in} \includegraphics{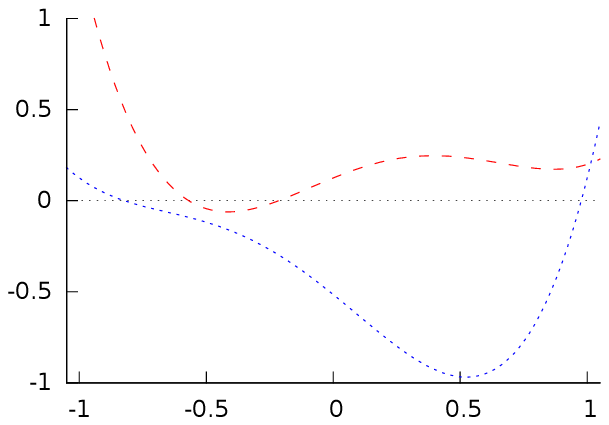} &
\vspace{-0.15in} \includegraphics{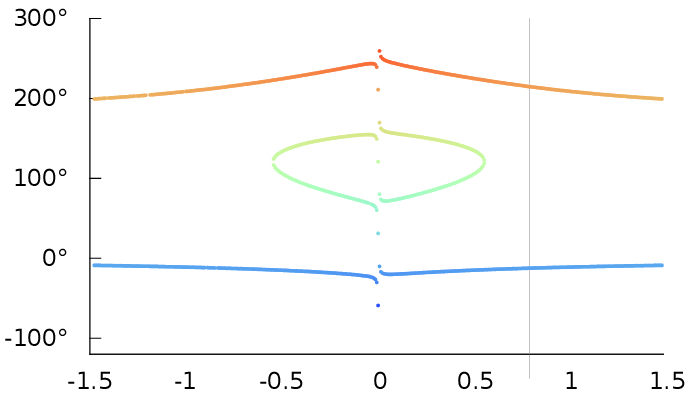} &
\vspace{0.1in} \includegraphics[scale=0.5]{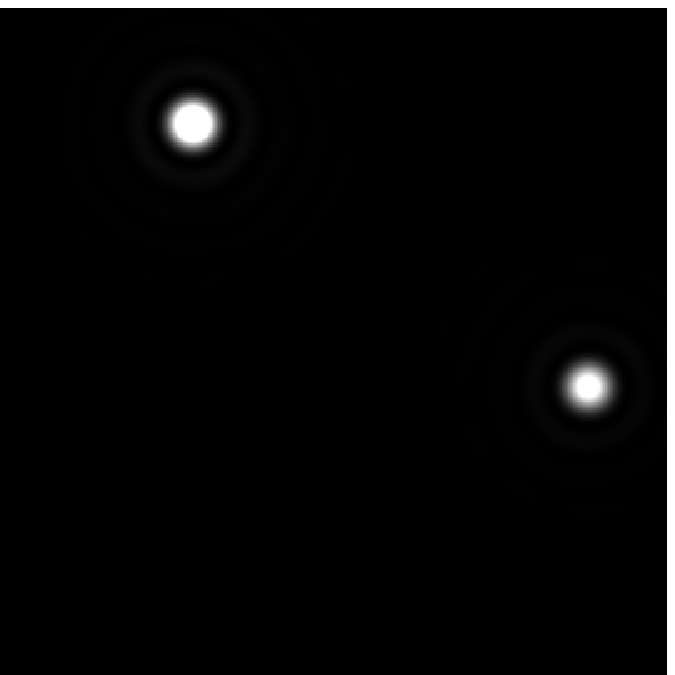}
\end{tabular}
\caption{\label{fig:diagroots}The quartic polynomials for $x=\sin(\phi_\xi-\phi)$ (dashed red line) and $y=\cos(\phi_\xi-\phi)$ (blue dotted line; see definitions in Eq.~(\ref{eq:zer01})), given by the left-hand sides of Eqs.~(\ref{eq:zer03*}) and (\ref{eq:zer0as0copy}), respectively. For this plot, the intermediate angle $\tfrac{1}{2}\mu=\pi/8$ was chosen, with $\nu=\{0,0.267,0.533,0.800\}$, corresponding to moving the telescope in increments of 1.5~m in this direction from the optical axis under the same parametrization as Fig.~\ref{fig:cross-0-3}. Noting the third row in particular, with the telescope near the caustic boundary, we observe how the near degeneracy of the telescopic view corresponds to the nearly degenerate roots of the cosine polynomial.}
\end{figure}

The final step in our derivation is the evaluation of the integral (\ref{eq:amp-dsf}) using the method of stationary phase, given the previously obtained values of $\phi_\xi$ from the quartic solution. Using the definition for the phase $\varphi$ of that integral and its second derivative  $\varphi''={d^2 \varphi}/{d\phi^2_\xi}$ given by (\ref{eq:zer*1*}) and (\ref{eq:phi-der2}), correspondingly, then recognizing that we have four solutions for $\phi^{[n]}_\xi, n\in[1,4]$, we obtain $\varphi_{[n]}=\varphi (\phi_{\xi[n]})=-\big(\alpha\rho\cos(\phi^{[n]}_\xi-\phi)+\beta_2\cos[2(\phi^{[n]}_\xi-\phi_s)]+{\textstyle\frac{\pi}{4}}\big)$ and $\varphi''_{[n]}=\varphi'' (\phi_{\xi[n]})$, leading to the result for intensity of light defined by (\ref{eq:S_mu}) that now is given as
 {}
\begin{eqnarray}
I({\vec x},{\vec x}_i)&=&
\bigg( \frac{1}{{2\pi}}\sum_{n=1}^4 \sqrt{\frac{2\pi}{|\varphi''_{[n]}|}}\Big(\frac{2J_1\big(u(\phi^{[n]}_\xi,\phi_i){\textstyle\frac{1}{2}}d\big)}{u(\phi^{[n]}_\xi,\phi_i){\textstyle\frac{1}{2}}d}\Big)
\cos\Big[\alpha\rho\cos(\phi^{[n]}_\xi-\phi)+\beta_2\cos[2(\phi^{[n]}_\xi-\phi_s)]+{\rm sign}[\varphi''_{[n]}] {\textstyle\frac{\pi}{4}}\Big]\bigg)^2 +\nonumber\\
 &+&
\bigg( \frac{1}{{2\pi}}\sum_{n=1}^4 \sqrt{\frac{2\pi}{|\varphi''_{[n]}|}} \Big(\frac{2J_1\big(u(\phi^{[n]}_\xi,\phi_i){\textstyle\frac{1}{2}}d\big)}{u(\phi^{[n]}_\xi,\phi_i){\textstyle\frac{1}{2}}d}\Big)
\sin\Big[\alpha\rho\cos(\phi^{[n]}_\xi-\phi)+\beta_2\cos[2(\phi^{[n]}_\xi-\phi_s)]+{\rm sign}[\varphi''_{[n]}] {\textstyle\frac{\pi}{4}}\Big]\bigg)^2,~~~
 \label{eq:Pvdfd0}
\end{eqnarray}
where
{}
\begin{eqnarray}
u(\phi^{[n]}_\xi,\phi_i)&=&\sqrt{\alpha^2+2\alpha\eta_i\cos(\phi^{[n]}_\xi-\phi_i)+\eta_i^2},
  \label{eq:eps343s}
\end{eqnarray}
also
$\phi^{[n]}_\xi=\phi^{[n]}_\xi(\bar u,\phi_u,\phi_s)$ are the real quartic solutions (up to four) of the equation
{}
\begin{eqnarray}
\alpha \rho\sin(\phi^{[n]}_\xi-\phi)+2\beta_2\sin[2(\phi^{[n]}_\xi-\phi_s)=0,
  \label{eq:int-ext-78*}
\end{eqnarray}
and the quantities $u(\phi^{[n]}_\xi,\phi_i)$, $\varphi_{[n]}$ and $\varphi''_{[n]}$ are   given as
{}
\begin{eqnarray}
\varphi_{[n]}&=&\alpha\rho\cos(\phi^{[n]}_\xi-\phi)+\beta_2\cos[2(\phi^{[n]}_\xi-\phi_s)]+{\rm sign}[\varphi''_{[n]}]{\textstyle\frac{\pi}{4}},\nonumber\\
\varphi''_{[n]}&=&-\big(\alpha\rho\cos(\phi^{[n]}_\xi-\phi)+4\beta_2\cos[2(\phi^{[n]}_\xi-\phi_s)]\big).
  \label{eq:eps34s}
\end{eqnarray}

These expressions describe the intensity in the focal plane of an imaging telescope.

\subsection{Interpreting the results}

To make sense of these results, we first note that the expression $2J_1(x)/x$ is maximal when $x=0$. Inspecting the argument of $J_1$ in (\ref{eq:Pvdfd0}), given in the form of $u(\phi^{[n]}_\xi,\phi_i)$ by (\ref{eq:eps343s}) we note that it is zero if and only if
\begin{align}
\eta_i&=\alpha, \qquad
\phi_i=\phi_\xi^{[n]}.
\end{align}
This immediately reveals the meaning of the solution that we obtained. The value $\eta_i=\alpha$ defines the radial position where the Einstein ring is formed in the focal plane of the telescope. The values of $\phi_\xi^{[n]}$ define the four spots of light that constitute the Einstein-cross, which appears as a result of a quadrupole gravitational lens.

To see that this is indeed the case, note Fig.~\ref{fig:cross-0-3}. This figure depicts the result for various telescope positions, as the telescope is moved from the center of the PSF towards one of its cusps. Multiple cases are shown in rows to reveal the gradual transition as the telescope reaches the caustic region, as well as subtle differences between telescopic images modeled using the quartic equation vs. computed through numerical integration. The first column in the figure offers a cross-sectional view of the quartic roots in the $\mu=0$ direction, parameterized on the horizontal axis by $\eta$. The telescope position is indicated by a thin vertical line. The next column shows an image of the PSF with the telescope position indicated by crosshairs. The third column shows the resulting telescope image, computed using the quartic solution. Finally, the fourth column shows the same simulated telescopic image using direct numerical integration.

When the telescope is not anywhere near the caustic boundary, the quartic solution, convolved with the PSF of the thin lens telescope, accurately reproduces the view seen through that telescope. However, as the telescope approaches the caustic boundary, one of the cusps of the PSF in this case, the situation changes. Numerical integration reveals that one of the spots begins to widen, forming a partial Einstein arc. The quartic solution cannot reproduce this feature, since it picks a specific eikonal angle, not a range of angles that correspond to an arc. Therefore, the quartic solution only shows a brightening of the corresponding spot, but the shape of the spot is unchanged.

This series of images is continued in Fig.~\ref{fig:cross-4-7}. As the leftmost column shows, we are approaching the caustic boundary where two of the four roots of the quartic equation vanish. As the roots converge in value, the corresponding spots brighten and approach each other. In contrast, numerical integration shows that the spots actually merge into a continuous arc.

In the next to last row in Fig.~\ref{fig:cross-4-7}, the telescope is now just outside the caustic boundary, in a region where the quartic equation only has two roots. The other two roots vanish suddenly; the spot corresponding to the remaining root shows considerable brightening (overexposed in this image with the concentric pattern of the $J_1$ Bessel function prominently visible) whereas numerical integration still shows a distinct arc. This process continues in the final image in the last row: the intensity of the remaining spot diminishes, whereas numerical integration shows the arc contracting to a bright spot, the minimum size of which is ultimately determined by the angular resolution of the imaging telescope.

We also looked at the telescope moving in the direction of the fold, given by $\tfrac{1}{2}\mu=\pi/4$ (Fig.~\ref{fig:fold-1-4}).
Such a movement is even more revealing of the behavior of the quartic solution. As the telescope approaches the caustic boundary, numerical integration shows two of the spots widening into arcs; this is not seen in the quartic solution. Furthermore, when the telescope passes the caustic boundary, two spots in the quartic solution abruptly vanish. Numerical integration, in contrast, show the two spots merging into an arc, which diminishes rapidly, but not instantaneously.

Yet another way of analyzing the roots comes from plotting the polynomials themselves. For special angles like $\tfrac{1}{2}\mu=0$ or $\tfrac{1}{2}\mu=\pi/4$, the roots are degenerate. For this reason, we picked an intermediate angle, $\tfrac{1}{2}\mu=\pi/8$, for Fig.~\ref{fig:diagroots}, which shows the telescope being moved in this direction and the resulting changing behavior of the polynomials on the left-hand sides of (\ref{eq:zer03*}) and (\ref{eq:zer0as0copy}). We can see two of the roots of both polynomials converge and vanish as the telescope approaches the caustic boundary of the PSF. Beyond the boundary, the remaining roots settle on their asymptotic behavior: In this case, the two roots of (\ref{eq:zer03*}) converge towards 0, whereas the two roots of (\ref{eq:zer0as0copy}) converge towards $\pm 1$, indicating that ultimately, we end up with two spots of light that are $180^\circ$ apart, corresponding to the telescopic image produced by the monopole lens.

These simulations offer valuable lessons as we explore the domain of applicability of the quartic solution, while keeping in mind that as it involves only direct computation and requires no numerical integration or other approximation techniques, it is a very powerful method of numerical modeling of quadrupole gravitational lenses.

\subsection{Numerical tools}

Images, including the images of the PSF as well as simulated images seen through a thin lens telescope, were produced using a modular software suite that we developed for the purpose of modeling the SGL.

This modular implementation allows us to flexibly use either the quartic solution that is the subject of this paper, or direct numerical integration of (\ref{eq:B2}) or its quadrupole-only variant, (\ref{eq:zer*1}). Furthermore, the modular architecture allows us to convolve the gravitational lens PSF with any additional PSF of our choice: this can be the trivial, ``identity'' PSF or it can be the PSF representing the thin lens telescope, as it appears, e.g., in (\ref{eq:amp-dsf}).

The quartic solution presented in this paper allows us to rapidly explore the behavior of the lens, with special emphasis on the actual observable: the view seen by an imaging telescope. Our implementation is sufficiently efficient to be useful for producing image frames for animation, and also to explore the lens in the spectral domain, creating multiple images, animations even, using a variety of wavelengths.

For instance, as part of our ongoing effort to evaluate the feasibility to use the SGL for realistic imaging of exoplanets, we were able to rapidly explore the vast parameter space associated with the SGL, iterating through a multitude of values representing total integration time, telescope aperture, heliocentric distance of the spacecraft, target distance, solar latitude and pixel resolution. The quartic solution, in conjunction with Fourier-deconvolution methods, allowed us to compute the resulting signal-to-noise ratio in nearly 6,000 cases in a matter of hours on desktop-class hardware.

We stress that not only does this yield a great improvement in computationally representing the quadrupole lens and modeling the resulting Einstein-cross, but that the method presented in this paper and the associated algorithms offer a proper wave-theoretical description, and can thus also be readily (and rapidly!) used in the modeling of spectral observables. This is the subject of our ongoing effort, of which the present paper represents a critically important step.

Our software tools, written mostly in the C and C++ programming languages, are evolving as we continue to develop them, aiming to have a reliable modeling set to be used primarily for evaluating observational scenarios using the SGL. Once mature, the code will be shared with the interested scientific community through appropriate channels.

\section{Discussion and Conclusions}
\label{sec:end}

While studying lensing by a weakly aspherical gravitational lens, such as the Sun, we found an expression (\ref{eq:B2}) that describes the EM field on the image plane after the wave traveled in the vicinity of an extended axisymmetric gravitating body (see \cite{Turyshev-Toth:2021-multipoles}). The gravitational part of the overall phase shift in that expression has a distinct signature of the lens' extended nature given by the term with zonal harmonics $J_n$. Each of these harmonics yields a caustic with distinct properties observed in the PSF of the extended axisymmetric lens (see discussion in  \cite{Turyshev-Toth:2021-caustics,Turyshev-Toth:2021-imaging}.)

In general, the integral (\ref{eq:B2}) must be treated numerically. However, in the case when only the contributions of the monopole and quadrupole harmonics are present, the resulting lens equation (\ref{eq:zer*1}) may be solved analytically. In this paper we pursued that opportunity and  studied the astroid caustic formed by a quadrupole gravitational lens using algebraic methods. This case deserves a special attention as it may yield additional insight into lensing by aspherical gravitational fields. We were able to develop an approximate solution to the new type of the diffraction integral (\ref{eq:zer*1}) that characterizes diffraction of light by non-spherical compact lenses.  That solution was obtained using the method of stationary phase thus reducing the problem to solving a quartic equation.

Although solvable analytically, quartic equations are usually solved numerically, because of the complexity and numerical stability of the algebraic approach. Defying convention, we opted to use algebraic methods anyway, and this brought unexpected benefits, yielding a very powerful method of numerical modeling that works almost everywhere except for the very narrow region around the caustic boundary.

The quartic solution offers a very efficient tool to study the PSF that is formed in the image plane in the strong interference region of the lens.  The solution is valid everywhere in the image plane except for a very narrow region in the immediate vicinity surrounding the astroid caustic. Studying the algebraic roots of the quartic equation offered unanticipated insight into the nature of the quadrupole gravitational lens. The discriminant of the quartic equation was found to correspond to the caustic region of the quadrupole PSF. The sharp boundary of this region in the quartic solution, contrasted with the more gradual transition between the caustic interior and exterior that we see through direct integration, allowed us to understand the limits of the method of stationary phase in this scenario.

Going beyond the PSF, we were also able to use the quartic solution in conjunction with an optical telescope modeled as a thin lens, and study the resulting telescopic images. We found that the roots of the quartic equation, expressed as angles, directly identify the location of bright spots on the Einstein ring that form the famous Einstein cross of a quadrupole lens. Moreover, we were able to study changes of the Einstein cross as a result of the telescope's displacement with respect to the optical axis of a point source. Here, we could also clearly see the limitations of the method of stationary phase: as the telescope approaches the caustic boundary and the light spots of the Einstein cross widen or merge into arcs, the method of stationary phase in the angular direction proves inadequate, as it cannot represent such arcs. On the other hand, outside the narrow caustic boundary region the method offers spectacular agreement with results obtained through direct integration, both in the interior and also outside the caustic boundary.

Direct computation of the roots of the quartic equation is very efficient. The sole caveat arises from the nature of these solutions, namely that real roots are often obtained after additive cancelation of imaginary parts. Even small rounding errors can lead to unacceptably large errors after quantities of similar magnitude but opposite in sign are added. This can lead to unexpected numerical instabilities. Such instabilities can be avoided, or at least mitigated, by well-known numerical techniques, such as intentionally introducing small errors, e.g., by displacing the coordinate system used for calculations by a tiny, random amount in a random direction, and thus avoiding cases where exact cancelation matters.

Once these matters are taken care of, the quartic method that we presented in this paper can serve as a very efficient way to compute the PSF of any gravitational lens that is dominated by its quadrupole moment, in all regions except the vicinity of its caustic boundary. This will prove important in our continuing study of the realistic Sun that is used as a gravitational lens to image distant objects such as exoplanets in other solar systems. The quartic solution will, therefore, serve as a vital tool in our effort to model image convolution and deconvolution, assess the resulting signal-to-noise ratios and estimate integration time and navigational requirements for the successful use of the SGL as an imaging tool.
This work is a critical step in our ongoing study of the SGL as the means for multipixel, multispectral imaging of exoplanets \cite{Turyshev-Toth:2020-extend,Toth-Turyshev:2020}  in the context of a realistic space mission \cite{Turyshev-etal:2020-PhaseII}.

Interesting results were obtained regarding the formation of the Einstein cross on the image sensor of an imaging telescope. We observed that an imaging telescope placed inside the astroid caustic observes four bright spots, forming the well-known pattern of an Einstein cross. The relative intensities and positions of these spots change as the telescope is moved in the image plane, with spots merging into bright arcs when the telescope approaches the caustic boundary. Outside the astroid caustic, only two spots remain, and the observed pattern eventually becomes indistinguishable from the imaging pattern of a monopole lens at greater distances from the optical axis.  We explored the formation of the Einstein cross and presented an algebraic solution for it using the Cardano solution for a quartic equation -- the first practical application of this famous solution with its nearly 500 year old \cite{Cardano:1545} history.

Concluding, we emphasize that not only do our results provide a first look at the properties of the new diffraction integral (\ref{eq:B2}) in the case of quadrupole gravitational lenses (\ref{eq:zer*1}), the method may also be used to model realistic astrophysical lenses, including stars, globular clusters, and distant spiral and spheroidal galaxies, especially when dealing with  lensing on the external gravitational potentials of these extended objects. This work is ongoing and results will be reported as they become available.

\begin{acknowledgments}
This work in part was performed at the Jet Propulsion Laboratory, California Institute of Technology, under a contract with the National Aeronautics and Space Administration.
VTT acknowledges the generous support Plamen Vasilev and other Patreon patrons.
\end{acknowledgments}


\begin{thebibliography}{18}
\expandafter\ifx\csname natexlab\endcsname\relax\def\natexlab#1{#1}\fi
\expandafter\ifx\csname bibnamefont\endcsname\relax
  \def\bibnamefont#1{#1}\fi
\expandafter\ifx\csname bibfnamefont\endcsname\relax
  \def\bibfnamefont#1{#1}\fi
\expandafter\ifx\csname citenamefont\endcsname\relax
  \def\citenamefont#1{#1}\fi
\expandafter\ifx\csname url\endcsname\relax
  \def\url#1{\texttt{#1}}\fi
\expandafter\ifx\csname urlprefix\endcsname\relax\def\urlprefix{URL }\fi
\providecommand{\bibinfo}[2]{#2}
\providecommand{\eprint}[2][]{\url{#2}}

\bibitem[{\citenamefont{{Turyshev} and
  {Toth}}(2021{\natexlab{a}})}]{Turyshev-Toth:2021-multipoles}
\bibinfo{author}{\bibfnamefont{S.~G.} \bibnamefont{{Turyshev}}}
  \bibnamefont{and} \bibinfo{author}{\bibfnamefont{V.~T.}
  \bibnamefont{{Toth}}}, \bibinfo{journal}{Phys. Rev. D}
  \textbf{\bibinfo{volume}{103}}, \bibinfo{pages}{064076}
  (\bibinfo{year}{2021}{\natexlab{a}}), \bibinfo{note}{arXiv:2102.03891
  [gr-qc]}.

\bibitem[{\citenamefont{{Turyshev} and
  {Toth}}(2021{\natexlab{b}})}]{Turyshev-Toth:2021-caustics}
\bibinfo{author}{\bibfnamefont{S.~G.} \bibnamefont{{Turyshev}}}
  \bibnamefont{and} \bibinfo{author}{\bibfnamefont{V.~T.}
  \bibnamefont{{Toth}}}, \bibinfo{journal}{Phys. Rev. D}
  \textbf{\bibinfo{volume}{104}}, \bibinfo{pages}{024019}
  (\bibinfo{year}{2021}{\natexlab{b}}), \bibinfo{note}{arXiv:2103.06955
  [gr-qc]}.

\bibitem[{\citenamefont{{Turyshev} and
  {Toth}}(2021{\natexlab{c}})}]{Turyshev-Toth:2021-imaging}
\bibinfo{author}{\bibfnamefont{S.~G.} \bibnamefont{{Turyshev}}}
  \bibnamefont{and} \bibinfo{author}{\bibfnamefont{V.~T.}
  \bibnamefont{{Toth}}}, \bibinfo{journal}{Phys. Rev. D}
  \textbf{\bibinfo{volume}{104}}, \bibinfo{pages}{044032}
  (\bibinfo{year}{2021}{\natexlab{c}}), \bibinfo{note}{arXiv:2104.08442
  [gr-qc]}.

\bibitem[{\citenamefont{{Turyshev} and {Toth}}(2017)}]{Turyshev-Toth:2017}
\bibinfo{author}{\bibfnamefont{S.~G.} \bibnamefont{{Turyshev}}}
  \bibnamefont{and} \bibinfo{author}{\bibfnamefont{V.~T.}
  \bibnamefont{{Toth}}}, \bibinfo{journal}{Phys. Rev. D}
  \textbf{\bibinfo{volume}{96}}, \bibinfo{pages}{024008}
  (\bibinfo{year}{2017}), \eprint{arXiv:1704.06824 [gr-qc]}.

\bibitem[{\citenamefont{{Turyshev} and
  {Toth}}(2020{\natexlab{a}})}]{Turyshev-Toth:2020-extend}
\bibinfo{author}{\bibfnamefont{S.~G.} \bibnamefont{{Turyshev}}}
  \bibnamefont{and} \bibinfo{author}{\bibfnamefont{V.~T.}
  \bibnamefont{{Toth}}}, \bibinfo{journal}{Phys. Rev. D}
  \textbf{\bibinfo{volume}{102}}, \bibinfo{pages}{024038}
  (\bibinfo{year}{2020}{\natexlab{a}}), \bibinfo{note}{arXiv:2002.06492
  [astro-ph.IM]}.

\bibitem[{\citenamefont{{Schneider} et~al.}(1992)\citenamefont{{Schneider},
  {Ehlers}, and {Falco}}}]{Schneider-Ehlers-Falco:1992}
\bibinfo{author}{\bibfnamefont{P.~S.} \bibnamefont{{Schneider}}},
  \bibinfo{author}{\bibfnamefont{J.}~\bibnamefont{{Ehlers}}}, \bibnamefont{and}
  \bibinfo{author}{\bibfnamefont{E.}~\bibnamefont{{Falco}}},
  \emph{\bibinfo{title}{Gravitational Lenses}}
  (\bibinfo{publisher}{Springer-Verlag Berlin Heidelberg},
  \bibinfo{year}{1992}).

\bibitem[{\citenamefont{Turyshev and Toth}(2019)}]{Turyshev-Toth:2019}
\bibinfo{author}{\bibfnamefont{S.~G.} \bibnamefont{Turyshev}} \bibnamefont{and}
  \bibinfo{author}{\bibfnamefont{V.~T.} \bibnamefont{Toth}},
  \bibinfo{journal}{Phys. Rev. D} \textbf{\bibinfo{volume}{99}},
  \bibinfo{pages}{024044} (\bibinfo{year}{2019}), \eprint{arXiv:1810.06627
  [gr-qc]}.

\bibitem[{\citenamefont{{Abramowitz} and
  {Stegun}}(1965)}]{Abramovitz-Stegun:1965}
\bibinfo{author}{\bibfnamefont{M.}~\bibnamefont{{Abramowitz}}}
  \bibnamefont{and} \bibinfo{author}{\bibfnamefont{I.~A.}
  \bibnamefont{{Stegun}}}, \emph{\bibinfo{title}{Handbook of Mathematical
  Functions: with Formulas, Graphs, and Mathematical Tables.}}
  (\bibinfo{publisher}{Dover Publications, New York; revised edition},
  \bibinfo{year}{1965}).

\bibitem[{\citenamefont{{Berry} and {Howls}}(2010)}]{Berry-Howles:2010}
\bibinfo{author}{\bibfnamefont{M.~V.} \bibnamefont{{Berry}}} \bibnamefont{and}
  \bibinfo{author}{\bibfnamefont{C.~J.} \bibnamefont{{Howls}}}, in
  \emph{\bibinfo{booktitle}{NIST Handbook of Mathematical Functions}}, edited
  by \bibinfo{editor}{\bibfnamefont{F.~W.} \bibnamefont{{Olver}}},
  \bibinfo{editor}{\bibfnamefont{D.~W.} \bibnamefont{{Lozier}}},
  \bibinfo{editor}{\bibfnamefont{R.~F.} \bibnamefont{{Boisvert}}},
  \bibnamefont{and} \bibinfo{editor}{\bibfnamefont{C.~W.}
  \bibnamefont{{Clark}}} (\bibinfo{publisher}{Cambridge University Press},
  \bibinfo{address}{Cambridge, UK}, \bibinfo{year}{2010}), pp.
  \bibinfo{pages}{775--793}.

\bibitem[{\citenamefont{{Park} et~al.}(2017)\citenamefont{{Park}, {Folkner},
  {Konopliv}, {Williams}, {Smith}, and {Zuber}}}]{Park-etal:2017}
\bibinfo{author}{\bibfnamefont{R.~S.} \bibnamefont{{Park}}},
  \bibinfo{author}{\bibfnamefont{W.~M.} \bibnamefont{{Folkner}}},
  \bibinfo{author}{\bibfnamefont{A.~S.} \bibnamefont{{Konopliv}}},
  \bibinfo{author}{\bibfnamefont{J.~G.} \bibnamefont{{Williams}}},
  \bibinfo{author}{\bibfnamefont{D.~E.} \bibnamefont{{Smith}}},
  \bibnamefont{and} \bibinfo{author}{\bibfnamefont{M.~T.}
  \bibnamefont{{Zuber}}}, \bibinfo{journal}{Astron. J.}
  \textbf{\bibinfo{volume}{153}}, \bibinfo{pages}{121} (\bibinfo{year}{2017}).

\bibitem[{\citenamefont{{Cardanus}}(1545)}]{Cardano:1545}
\bibinfo{author}{\bibfnamefont{H.}~\bibnamefont{{Cardanus}}},
  \emph{\bibinfo{title}{Ars magna}} (\bibinfo{year}{1545}),
  \bibinfo{note}{{English translation}: {\em The Rules of Algebra (Ars Magna)},
  trans: {{Witmer}, T. Richard}, Dover, 1993}.

\bibitem[{\citenamefont{{Garling}}(1986)}]{Garling:1986}
\bibinfo{author}{\bibfnamefont{D.~J.~H.} \bibnamefont{{Garling}}},
  \emph{\bibinfo{title}{A Course in Galois Theory}}
  (\bibinfo{publisher}{Cambridge University Press},
  \bibinfo{address}{Cambridge, UK}, \bibinfo{year}{1986}).

\bibitem[{\citenamefont{{Stewart}}(2004)}]{Stewart:2004}
\bibinfo{author}{\bibfnamefont{I.}~\bibnamefont{{Stewart}}},
  \emph{\bibinfo{title}{Galois Theory, Third Edition.}}
  (\bibinfo{publisher}{Chapman \& Hall/CRC Mathematics, CRC Press},
  \bibinfo{address}{Boca Raton, FL}, \bibinfo{year}{2004}).

\bibitem[{\citenamefont{{Szele}}(1975)}]{Szele:1975}
\bibinfo{author}{\bibfnamefont{T.}~\bibnamefont{{Szele}}},
  \emph{\bibinfo{title}{{Bevezet{\'e}s az algebr{\'a}ba (Introduction to
  algebra -- in Hungarian)}}} (\bibinfo{publisher}{{Tank{\"o}nyvkiad{\'o}}},
  \bibinfo{address}{Budapest}, \bibinfo{year}{1975}).

\bibitem[{\citenamefont{{Turyshev} and
  {Toth}}(2020{\natexlab{b}})}]{Turyshev-Toth:2020-image}
\bibinfo{author}{\bibfnamefont{S.~G.} \bibnamefont{{Turyshev}}}
  \bibnamefont{and} \bibinfo{author}{\bibfnamefont{V.~T.}
  \bibnamefont{{Toth}}}, \bibinfo{journal}{Phys. Rev. D}
  \textbf{\bibinfo{volume}{101}}, \bibinfo{pages}{044048}
  (\bibinfo{year}{2020}{\natexlab{b}}), \bibinfo{note}{arXiv:1911.03260
  [gr-qc]}.

\bibitem[{\citenamefont{{Toth} and {Turyshev}}(2021)}]{Toth-Turyshev:2020}
\bibinfo{author}{\bibfnamefont{V.~T.} \bibnamefont{{Toth}}} \bibnamefont{and}
  \bibinfo{author}{\bibfnamefont{S.~G.} \bibnamefont{{Turyshev}}},
  \bibinfo{journal}{Phys. Rev. D} \textbf{\bibinfo{volume}{103}},
  \bibinfo{pages}{124038} (\bibinfo{year}{2021}),
  \bibinfo{note}{arXiv:2012.05477 [gr-qc]}.

\bibitem[{\citenamefont{{Turyshev} et~al.}(2020)\citenamefont{{Turyshev},
  {Shao}, {Toth}, and et~al.}}]{Turyshev-etal:2020-PhaseII}
\bibinfo{author}{\bibfnamefont{S.~G.} \bibnamefont{{Turyshev}}},
  \bibinfo{author}{\bibfnamefont{M.}~\bibnamefont{{Shao}}},
  \bibinfo{author}{\bibfnamefont{V.~T.} \bibnamefont{{Toth}}},
  \bibnamefont{and} \bibinfo{author}{\bibnamefont{et~al.}},
  \emph{\bibinfo{title}{Direct multipixel imaging and spectroscopy of an
  exoplanet with a solar gravity lens mission}} (\bibinfo{year}{2020}),
  \bibinfo{note}{arXiv:2002.11871 [gr-qc]}.

\bibitem[{\citenamefont{{Turyshev} and
  {Toth}}(2019)}]{Turyshev-Toth:2019-fin-difract}
\bibinfo{author}{\bibfnamefont{S.~G.} \bibnamefont{{Turyshev}}}
  \bibnamefont{and} \bibinfo{author}{\bibfnamefont{V.~T.}
  \bibnamefont{{Toth}}}, \bibinfo{journal}{Phys. Rev. D}
  \textbf{\bibinfo{volume}{100}}, \bibinfo{pages}{084018}
  (\bibinfo{year}{2019}), \bibinfo{note}{arXiv:1908.01948 [gr-qc]}.

\end{thebibliography}

\appendix

\section{Studying two special cases:  $\alpha\rho\gg\beta_2$ and  $\beta_2\gg\alpha\rho$.}
\label{sec:mono-quad-cases}

\subsection{The case when $\alpha\rho\gg\beta_2$}
\label{sec:mono-case}

In the case when $\alpha\rho\gg\beta_2$ (or, equivalently, for very large
deviations from the optical axis, $\rho\gg\beta_2/\alpha$), we may neglect the contribution of the quadrupole. This effectively  reduces the problem to the situation when only monopole is present, while $J_2=0$.
In this caes, (\ref{eq:zer*1})  takes the form
{}
\begin{eqnarray}
B_0(\rho, \phi)&=&\frac{1}{{2\pi}}\int_0^{2\pi} d\phi_\xi \exp\Big[-i\alpha\rho\cos(\phi_\xi-\phi)\Big].
\label{eq:mon*1}
\end{eqnarray}

This integral can be evaluated directly (with $J_0$ is  the Bessel function of the first kind), yielding
{}
\begin{eqnarray}
B_0(\rho, \phi)&=&J_0(\alpha\rho),
\label{eq:mon*10}
\end{eqnarray}
which results in the following expression for the light amplification of a monopole SGL \cite{Turyshev-Toth:2017}
 {}
\begin{eqnarray}
\mu_0({\vec x})={2\pi kr_g}  \, J^2_0(\alpha\rho).
  \label{eq:S_mu0}
\end{eqnarray}

However, it is instructive to evaluate the integral (\ref{eq:mon*1}) with the method of the stationary phase and  to consider the differences in the final result, if any. To implement this objective, we identify the phase of the integrand in (\ref{eq:mon*1}):
{}
\begin{eqnarray}
\varphi(\rho, \phi)&=&-\alpha\rho\cos(\phi_\xi-\phi).
\label{eq:mon*1=*}
\end{eqnarray}
Next, we take the first and second derivatives of the pahse, $\varphi'=d\varphi/d\phi_\xi$ and  $\varphi''=d^2\varphi/d\phi_\xi^2$,  correspondingly, yielding
{}
\begin{eqnarray}
\varphi'=
\frac{d\varphi}{d\phi_\xi}&=&\alpha\rho\sin(\phi_\xi-\phi),
\label{eq:phi-der-mon}\\
\varphi''=
\frac{d^2\varphi}{d\phi_\xi^2}&=&\alpha\rho\cos(\phi_\xi-\phi).
\label{eq:phi-der2-mon}
\end{eqnarray}

The phase is stationary, when $\varphi'=0$, yielding two solutions for the angle $\phi_\xi$, namely $(\phi_\xi-\phi)=\{0,\pi\}$ and, thus, $\varphi''=\pm\alpha\rho$. In terms of the values of $(\varphi_n, \varphi''_n)$ from (\ref{eq:phi-der-mon}) and (\ref{eq:phi-der2-mon}), we have two solutions: $(\varphi_1, \varphi''_1)=(-\alpha\rho,\alpha\rho)$ and $(\varphi_2, \varphi''_2)=(\alpha\rho,-\alpha\rho)$. Then, the solution for  (\ref{eq:mon*1}) takes the form
{}
\begin{eqnarray}
B_0(\rho, \phi)&=&\frac{1}{{2\pi}}\int_0^{2\pi} d\phi_\xi \exp\Big[-i\alpha\rho\cos(\phi_\xi-\phi)\Big]=\frac{1}{{2\pi}}\sum_{n=1,2}\Big(\frac{2\pi}{|\varphi''_n|}\Big)^{1/2}e^{i\big(\varphi_{0n}+{\rm sign}[\varphi''_n]{\textstyle\frac{\pi}{4}}\big)}=\nonumber\\
&=&\frac{1}{\sqrt{2\pi}}\frac{1}{\sqrt{\alpha\rho}}\Big(e^{i(\alpha\rho-{\textstyle\frac{\pi}{4}})}+e^{-i(\alpha\rho-{\textstyle\frac{\pi}{4}})}\Big)=
\sqrt{\frac{2}{\pi\alpha\rho}}\cos(\alpha-{\textstyle\frac{\pi}{4}}),~~~~~
\label{eq:mon*1s}
\end{eqnarray}
which is the approximation for Bessel function $J_0(\alpha\rho)$ from (\ref{eq:mon*10}) for the large values of the argument, $\alpha\rho\gg1$.

Next, we compute the light amplification of the SGL given by the following expression
 {}
\begin{eqnarray}
\mu_z({\vec x})={2\pi kr_g}  \, B^2_0(\rho, \phi).
  \label{eq:S_muz}
\end{eqnarray}
With this, the SGL's magnification factor in the case of a monopole takes the familiar form \cite{Turyshev-Toth:2019-fin-difract}:
{}
\begin{eqnarray}
\mu_{z0}&=&2\pi k r_gB^2_0(\rho,\phi)=\frac{4k r_g}{\alpha\rho}\cos^2(\alpha\rho-{\textstyle\frac{\pi}{4}})=
\frac{2\sqrt{2r_gr}}{\rho}\cos^2\Big(k\sqrt\frac{2r_g}{r}\rho-{\textstyle\frac{\pi}{4}}\Big).
\label{eq:mu_0}
\end{eqnarray}

Comparing results (\ref{eq:mon*1s}) and (\ref{eq:mu_0}), with (\ref{eq:mon*10}) and (\ref{eq:S_mu0}), we see that the method of stationary phase produces results that are valid for $\alpha\rho\gg1$ and do not cover the region in the immediate vicinity of the optical axis where $0\leq\alpha\rho\ll1$, which is the highly-oscillating region.  It is important to recognize this limiting feature of the method of stationary phase that may not be used to treat the regions with high and rapidly growing oscillations
(i.e., near the caustics), but may provide good approximation everywhere else.

\subsection{The case when $\beta_2\gg\alpha\rho$.}
\label{sec:quad-case}

The case when $\beta_2\gg\alpha\rho$ (or, when the deviations from the optical axis are small, $\rho\ll\beta_2/\alpha$) we may neglect the presence of the $\alpha\rho$-term and treat the case with quadrupole only, $J_2\not=0$ and $\alpha\rho=0$. This case describes the interior region of the astroid caustic where the integral (\ref{eq:zer*1})  may be approximated taking the form
{}
\begin{eqnarray}
B_2(\rho, \phi)&=&\frac{1}{{2\pi}}\int_0^{2\pi} d\phi_\xi \exp\Big[-i\beta_2\cos[2(\phi_\xi-\phi_s)]\Big].
\label{eq:quad*1}
\end{eqnarray}
Clearly, this integral is yet another Bessel $J_0$ function (see Appendix~A in \cite{Turyshev-Toth:2021-caustics} for a more formal treatment of this integral in the vicinity of the optical axis, using Jacobi-Anger expansion \cite{Abramovitz-Stegun:1965}):
{}
\begin{eqnarray}
B_2(\rho, \phi)&=&J_0(\beta_2),
\label{eq:quad*10}
\end{eqnarray}
which yields expression for the light amplification of the quadrupolar SGL, valid in the region near the optical axis
 {}
\begin{eqnarray}
\mu_2({\vec x})={2\pi kr_g}  \, J^2_0(\beta_2).
  \label{eq:S_mu2}
\end{eqnarray}

Again, it is instructive to evaluate the integral by using the methods of the stationary phase. To implement this objective, we recognize that the phase of the integrand in  (\ref{eq:quad*1}), $\varphi_2$, has the form:
{}
\begin{eqnarray}
\varphi_2(\rho, \phi)&=&-\beta_2\cos[2(\phi_\xi-\phi_s)].
\label{eq:quad*1*}
\end{eqnarray}

To proceed, we take the derivatives $\varphi'_2=d\varphi_2/d\phi_\xi$ and  $\varphi''_2=d^2\varphi_2/d\phi_\xi^2$,  that yeilds
{}
\begin{eqnarray}
\varphi'_2=
\frac{d\varphi_2}{d\phi_\xi}&=&2\beta_2\sin[2(\phi_\xi-\phi_s)],
\label{eq:phi-der-quad}\\
\varphi''_2=
\frac{d^2\varphi_2}{d\phi_\xi^2}&=&4\beta_2\cos[2(\phi_\xi-\phi_s)].
\label{eq:phi-der2-quad}
\end{eqnarray}

The phase is stationary, when $\varphi'_2=0$, yielding two solutions $2(\phi_\xi-\phi_s)=\{0,\pi\}$ and, thus, $\varphi''_2=\pm4\beta_2$. As a result, we have four solutions with the values $(\varphi_n, \varphi''_n)$ were identified to be $(\varphi_1, \varphi''_1)=(\beta_2,-4\beta_2)$, $(\varphi_2, \varphi''_2)=(\beta_2,-4\beta_2)$, $(\varphi_3, \varphi''_3)=(-\beta_2,4\beta_2)$ and $(\varphi_4, \varphi''_4)=(-\beta_2,4\beta_2)$. Clearly, these are just two identical sets of values. Then, the solution for  (\ref{eq:quad*1}) takes the form
{}
\begin{eqnarray}
B_2(\rho, \phi)&=&\frac{1}{{2\pi}}\int_0^{2\pi} d\phi_\xi \exp\Big[-i\beta_2\cos[2(\phi_\xi-\phi_s)]\Big]=\frac{1}{{2\pi}}\sum_{n=1,2}\Big(\frac{2\pi}{|\varphi''_n|}\Big)^{1/2}e^{i\big(\varphi_{0n}+{\rm sign}[\varphi''_n]{\textstyle\frac{\pi}{4}}\big)}=\nonumber\\
&=&\frac{1}{\sqrt{2\pi}}\frac{2}{\sqrt{4\beta_2}}\Big(e^{i(\beta_2-{\textstyle\frac{\pi}{4}})}+e^{-i(\beta_2-{\textstyle\frac{\pi}{4}})}\Big)=
\sqrt{\frac{2}{\pi\beta_2}}\cos(\beta_2-{\textstyle\frac{\pi}{4}}),~~~~~
\label{eq:quad*1=}
\end{eqnarray}
where we recognize the approximation for the Bessel function $J_0(\beta_2)$ for large argument, $\beta_2\gg1$.

Now we can compute the SGL's magnification factor in the case of a quadrupole ($\beta_s\not=0$) that takes the form:
{}
\begin{eqnarray}
\mu_{z2}&=&2\pi k r_gB^2_2(\rho,\phi)=\frac{4k r_g}{\beta_2}\cos^2(\beta_2-{\textstyle\frac{\pi}{4}})=
\frac{2}{J_2\sin^2\beta_s}\Big(\frac{2r_gr}{R_\odot^2}\Big)\cos^2\Big(kJ_2 \frac{R_\odot^2}{2r}\sin^2\beta_s-{\textstyle\frac{\pi}{4}}\Big),
\label{eq:mu_0_quad}
\end{eqnarray}
which is independent on $\rho$ (as $\beta_2$ does not depend on $\rho$). This describes the case when $\beta_2\gg\alpha\rho$ in (\ref{eq:zer*1}) or for very small deviations from the optical axis, $\rho \ll\beta_2/\alpha$,  in the inner region of the astroid caustic.

Similar to the conclusion in Sec.~\ref{sec:mono-case}, we see that that method of the stationary phase produces an approximate result in the region of interest.  This fact should be kept in mind in the relevant analyzes.

\section{Solving for $\cos(\phi_\xi-\phi)$}
\label{sec:cos}

Going back to (\ref{eq:zer0}), we may solve it for $\cos(\phi_\xi-\phi)$.
For that, we use (\ref{eq:zer0}) and, similarly to (\ref{eq:zer_01+}), re-write it as
{}
\begin{eqnarray}
\sin(\phi_\xi-\phi)\Big(\alpha\rho+4\beta_2 \cos(\phi_\xi-\phi)\cos\mu\Big)+
2\beta_2\big(2\cos^2(\phi_\xi-\phi)-1\big)\sin\mu=0.
\label{eq:zer_01+d}
\end{eqnarray}

To solve (\ref{eq:zer_01+d}),  we define
{}
\begin{eqnarray}
\cos(\phi_\xi-\phi)=y, ~~~~ \sin(\phi_\xi-\phi)=\pm\sqrt{1-y^2}=\pm
x,
\label{eq:zeyr01}
\end{eqnarray}
where $x$ is given by (\ref{eq:zer01}) and, thus, $x^2+y^2=1$. With this definition of $y$, we re-write (\ref{eq:zer_01+d}) as below:
{}
\begin{eqnarray}
2\beta_2 (2y^2-1)\sin\mu\pm\Big(\alpha\rho +4\beta_2 y \cos\mu\Big)\sqrt{1-y^2}=0.
\label{eq:zeyr01=}
\end{eqnarray}

By isolating the term with the square root in Eq.~(\ref{eq:zeyr01=}) and squaring the result, we re-write (\ref{eq:zeyr01=}) as below:
{}
\begin{eqnarray}
16\beta^2_2\, y^4+8\alpha\rho\beta_2 \cos\mu \, y^3+\Big((\alpha\rho)^2-16\beta^2_2\Big)y^2-8\alpha\rho\beta_2\cos\mu \, y+4\beta^2_2\sin^2\mu-(\alpha\rho)^2=0.
\label{eq:zer01=d0}
\end{eqnarray}

If $\beta_2=0$, equation (\ref{eq:zer01=d0}) yields $y=\pm1$ or $\cos(\phi_\xi-\phi)=\pm1$, resulting in $\phi_\xi=\phi\pm{\textstyle\frac{1}{2}}\pi$, which is consistent with the fact that in in the case of a
gravitational monopole, the light rays propagate in a plane (see discussion in Appendix~\ref{sec:mono-case}).

If $\beta_2\not=0$, we divide (\ref{eq:zer01=d0}) by $16\beta^2_2$ and, using definition for $\eta$ from (\ref{eq:zer03}), we transform that equation as below:
{}
\begin{eqnarray}
y^4+2\eta \cos\mu \, y^3+(\eta^2-1)y^2-2\eta\cos\mu \, y+{\textstyle\frac{1}{4}}\sin^2\mu-\eta^2=0.
\label{eq:zer0as0}
\end{eqnarray}

This is our second key equation. It may be used to determine the appropriate signs in (\ref{eq:zer01}) to remove the $\pi$ ambiguity.

Using the generic solution (\ref{eq:zer04})--(\ref{eq:zer07*}), we present solution to (\ref{eq:zer0as0}) as below:
{}
\begin{eqnarray}
y_{1,2}&=&-{\frac{1}{2}}\eta\cos\mu-S_c\pm\frac{1}{2}\sqrt{-4S^2_c-2(\eta^2-1)+3\eta^2\cos^2\mu-\frac{\eta\cos\mu(1+\eta^2\sin^2\mu)}{S_c}},
\label{eq:sol01c}\\
y_{3,4}&=&-{\frac{1}{2}}\eta\cos\mu+S_c\pm\frac{1}{2}\sqrt{-4S^2_c-2(\eta^2-1)+3\eta^2\cos^2\mu+\frac{\eta\cos\mu(1+\eta^2\sin^2\mu)}{S_c}},
\label{eq:sol01bc}
\end{eqnarray}
where $S_c$ and $Q_c$ are given as below:
{}
\begin{eqnarray}
S_c&=&\frac{1}{2}\sqrt{-\frac{2}{3}(\eta^2-1)+\eta^2\cos^2\mu+\frac{1}{3}\Big(Q_c+\frac{(\eta^2-1)^2-3(4\eta^2-1)\sin^2\mu}{Q_c}\Big)},\\
Q_c&=&\Big\{(\eta^2-1)^3+18\eta^2(\eta^2+2)\cos^2\mu+18\Big({\textstyle\frac{1}{4}}\sin^2\mu-\eta^2\Big)\Big(3\eta^2\cos^2\mu-2(\eta^2-1)\Big)+
\nonumber \\&&\hskip 43pt~+
\sqrt{\Big(\cos^2\mu-2\eta^2\Big)^2\Big(27(\eta^2-1)^3+\Big(\frac{27}{2}\Big)^2\eta^4\sin^2\mu\Big)\sin^2\mu}\Big\}^{1/3}.~~~~~
\label{eq:sol02c}
\end{eqnarray}

\end{document}